\documentclass[twocolumn,showpacs,nofootinbib]{revtex4-1}

\usepackage{epsf}
\usepackage{graphicx} 
\usepackage{amssymb}
\usepackage{dcolumn}
\usepackage{amsmath}
\usepackage{hyperref}

\def\apj{ApJ\,}                 
\def\apjl{ApJL\,}                
\def\apjs{ApJS\,}               
\def\mnras{MNRAS\,}             
\def\aap{A\&A\,}                
\def\physrep{Phys.~Rep.\,}   
\def\araa{ARA\&A\,}             

\def\thetaB{\mbox{\boldmath$\theta$}}
\def\thetazB{\mbox{\boldmath$\theta_0$}}

\def\eB{\mbox{\boldmath$e$}}
\def\cB{\mbox{\boldmath$c$}}
\def\ellB{\mbox{\boldmath$\ell$}}
\def\gB{\mbox{\boldmath$g$}}
\def\gammaB{\mbox{\boldmath$\gamma$}}
\def\pB{\mbox{\boldmath$p$}}
\def\CB{\mbox{\boldmath${\rm C}$}}
\def\DB{\mbox{\boldmath$d$}}
\def\muB{\mbox{\boldmath$\mu$}}
\def\lsim{~\rlap{$<$}{\lower 1.0ex\hbox{$\sim$}}}
\def\gsim{~\rlap{$>$}{\lower 1.0ex\hbox{$\sim$}}}

\begin{document}

\title{Cosmology constraints from the weak lensing peak counts\\ and the power spectrum in CFHTLenS}

\author{Jia Liu$^{1}$}
\email {jia@astro.columbia.edu}

\author{Andrea Petri$^{2}$}
\email {apetri@phys.columbia.edu}

\author{Zolt\'an Haiman$^{1,3}$}
\email {zoltan@astro.columbia.edu}

\author{Lam Hui$^{2,3}$}
\email {lhui@astro.columbia.edu}

\author{Jan M. Kratochvil$^{4}$}
\email {kratochvilj@ukzn.ac.za}

\author{Morgan May$^{5}$}
\email {may@bnl.gov}

\affiliation{ {$^1$ Department of Astronomy and Astrophysics, Columbia University, New York, NY 10027, USA}} 
\affiliation{ {$^2$ Department of Physics, Columbia University, New York, NY 10027, USA}} 
\affiliation{ {$^3$ Institute for Strings, Cosmology, and Astroparticle Physics (ISCAP), Columbia University, New York, NY 10027, USA}} 
\affiliation{ {$^4$ Astrophysics and Cosmology Research Unit, University of KwaZulu-Natal, Westville, Durban, 4000, South Africa}   }
\affiliation{ {$^5$ Physics Department, Brookhaven National Laboratory, Upton, NY 11973, USA}   }

\date{\today}

\begin{abstract}
Lensing peaks have been proposed as a useful statistic, containing
cosmological information from non-Gaussianities that is inaccessible
from traditional two-point statistics such as the power spectrum or
two-point correlation functions.  Here we examine constraints on
cosmological parameters from weak lensing peak counts, using the
publicly available data from the 154 deg$^2$ CFHTLenS survey.  We
utilize a new suite of ray-tracing N-body simulations on a grid of 91
cosmological models, covering broad ranges of the three parameters
$\Omega_m$, $\sigma_8$, and $w$, and replicating the Galaxy sky
positions, redshifts, and shape noise in the CFHTLenS observations. We
then build an emulator that interpolates the power spectrum and the
peak counts to an accuracy of $\leq 5\%$, and compute the likelihood
in the three-dimensional parameter space ($\Omega_m$, $\sigma_8$, $w$)
from both observables.  We find that constraints from peak counts are 
comparable to those from the power spectrum, and somewhat tighter when 
different smoothing scales are combined.
Neither observable can constrain $w$ without external data.
When the power spectrum and peak counts are combined, the area of the error
``banana'' in the ($\Omega_m$, $\sigma_8$) plane reduces by a factor of $\approx2$,
compared to using the power spectrum alone. For a flat $\Lambda$ cold dark
matter model, combining both statistics, we obtain the constraint
$\sigma_8(\Omega_m/0.27)^{0.63}=0.85\substack{+0.03 \\ -0.03}$.

\end{abstract}

\pacs{PACS codes: 98.80.-k, 95.36.+x, 95.30.Sf, 98.62.Sb}

\maketitle

\section{Introduction}\label{Introduction}

Weak gravitational lensing (WL) is one of the most promising
techniques to probe dark energy (DE) with improved precision in the
future (see recent reviews by
\cite{Refregier2003,Schneider2005,Hoekstra2008,Bartelmann2010,Weinberg2013}). By
statistically measuring the distortions in the shapes of background
galaxies, the matter density fluctuations at different redshifts can
be mapped, yielding constraints on the parameters of the background
cosmological model. Pioneering WL surveys, such as the Cosmic
Evolution Survey~(COSMOS, \cite{Schrabback2010}) and the
Canada-France-Hawaii Telescope Lensing Survey~ (CFHTLenS,
\cite{CFHTLS2012,Kilbinger2013}) have recently successfully
demonstrated the utility of this technique, yielding constraints on
the matter density $\Omega_m$ and fluctuation amplitude
$\sigma_8$ comparable with other existing methods, even with
relatively small sky coverage ($\sim 1$ and 154 deg$^2$,
respectively).

In this paper, we use the publicly available CFHTLenS data on $\approx
4.2$ million galaxies, combined with a suite of ray-tracing
simulations in 91 different cosmological models, to constrain the
cosmological parameters, $\Omega_m$, $\sigma_8$, and the DE
equation of state $w$. Traditionally, WL data is analyzed using the
two-point correlation function (2PCF), or its Fourier-space
counterpart, the power spectrum.  However, these statistics can not
fully characterize the weak lensing shear field on small ($\lsim$
arcmin) angular scales, where it is sensitive to matter density
fluctuations in the nonlinear regime, and is strongly non-Gaussian.
Various non-Gaussian statistics (e.g. higher moments~\cite{Bernardeau1997,
Hui1999, vanWaerbeke2001, Takada2002, Zaldarriaga2003, Kilbinger2005}, 
three-point functions~\cite{Takada2003, Vafaei2010}, 
bispectra~\cite{Takada2004,DZ05,Sefusatti2006,Berge2010}, 
peak counts~\cite{Jain2000b, Marian2009, Maturi2010, Yang2011,Marian+2013}, 
or Minkowski functionals~\cite{Kratochvil2012, Petri2013}) have been proposed in the past, 
and shown to improve cosmological constraints from WL surveys.

In this work, we focus on peak statistics, which describe the
distribution of local maxima in a convergence map, as a function of
peak height. It is a particularly simple statistic, forecasted to yield
a factor of $\sim$ two improvement on cosmological parameters 
when combined with two-point statistics by several recent studies
\cite{Dietrich2010,Kratochvil2010,Marian2011,Marian2012,Yang2013}, and
also found to be unusually robust to systematic errors from baryonic
effects \cite{Yang2013}.  In a companion paper (Petri et al. in prep)
we examine constraints from Minkowski functionals and higher moments
of the WL convergence field.

A handful of works have recently begun to examine non-Gaussian
features in the CFHTLenS data.  Three-point statistics have been
measured in both CFHTLenS \cite{Fu2014} and earlier in COSMOS
\cite{Semboloni2011b} and found to lead to modest (up to $\approx
10\%$) improvements on the combination $\sigma_8 \Omega_m^\alpha$ with
$\alpha \approx 0.3-0.5$.  Ref.~\cite{ShirasakiYoshida2014} measured
Minkowski functionals in CFHTLenS and showed that they can break
degeneracies among cosmological parameters, improving constraints on
$\Omega_m$ and $\sigma_8$. Finally, higher moments
\cite{VanWaerbeke2013} and peak counts \cite{Shan2012} have both been
measured in CFHTLenS , although cosmological constraints have not yet
been derived from them.

The distinguishing feature of the present work is that we compute peak
count statistics, including their dependence on cosmology and their
variance, from simulations in a large number of cosmological models
(91 in total).  Simulating multiple cosmological models is necessary because analytical 
predictions of peak counts for non-Gaussian fields are still in early 
development (for example, Ref.\cite{Fan2010}). Furthermore, a large number ($\gsim$ hundreds)
of realizations per model is necessary to measure the covariance of
the peak counts, and to compute accurate confidence limits on
cosmological parameters.  Because of computational limitations, most
works on non-Gaussian WL statistics to date have sampled only a few
points in the multi-dimensional cosmological parameter space, and
assumed a linear dependence on cosmological parameters to compute
observables in other cosmologies (effectively implementing a numerical
version of a Fisher matrix) or else relied on fitting formulae
calibrated with a handful of simulations.  The only exception we are
aware of is Ref.~\cite{Dietrich2010}, who studied peak counts in
simulations on a two-dimensional $\Omega_m,\sigma_8$ grid, and whose
results already indicate that the $\Omega_m$ and $\sigma_8$ dependence
is nonlinear, and the Fisher approach is therefore highly inaccurate.

Recently, a series of papers dubbed ``the Coyote Universe''
\cite{Heitmann2010I, Heitmann2009II, Heitmann2010III, Heitmann2014E}
have built an emulator, based on a large number of simulations, to
address analogous issues for the matter power spectrum. Using 37
cosmological models, these studies have shown that the matter power
spectrum can be interpolated and computed to $~1\%$ accuracy out to 
$k\sim1$ Mpc$^{-1}$ for
models in-between the simulated points in parameter space.  We have
built an emulator following a similar approach, but describing WL
observables, and tailored specifically for the CFHTLenS fields. Unlike
in a general--purpose emulator, galaxy properties (e.g. redshift
distribution, position, and noise) are not freely adjustable
parameters, but rather fixed and built into our simulations from the
outset, adapted directly from the CFHTLenS measurements. 

The paper is structured as follows. We first describe CFHTLenS data
processing and convergence map construction in \S~\ref{sec: data}, and our
ray-tracing simulations and numerical details in \S~\ref{sec:
  emulator}. We present the results of our analysis in \S~\ref{sec:
  results}, and we offer our conclusions in \S~\ref{sec: conclusion}.

\section{CFHTLenS Data Processing}\label{sec: data}

The 154 deg$^2$ CFHTLenS data cover four individual patches on the
sky, with an area of 64, 23, 44 and 23~deg$^2$ for field W1, W2, W3
and W4, respectively. The CFHTLenS data analysis roughly consists of: (1)
creation of the galaxy catalogue using SExtractor \cite{Erben2013};
(2) the photometric redshift estimation with a Bayesian photometric
redshift code \cite{Hildebrandt2012} ; (3) galaxy shape measurement
with {\it lens}fit \cite{Heymans2012, Miller2013}; and finally (4)
cosmological analysis with 2PCF~\cite{Kilbinger2013}. A summary of the
data analysis process is listed in Appendix C of
Ref.~\cite{Erben2013}. We refer the readers to the CFHTLenS papers
mentioned above for more technical details.

We apply the following cuts to galaxies: mask $\leq 1$ (see Table~B2 in 
Ref.~\cite{Erben2013} for the meaning of mask values), redshift
$0.2<z<1.3$, fitclass $= 0$ (requiring the object to be a galaxy), and
weight $\mathcal{W} > 0$ (with larger $\mathcal{W}$ indicating smaller shear measurement
uncertainty).  Applying these cuts leaves us 4.2 million galaxies,
124.7 deg$^2$ sky, and average number density $n_{gal}\approx
9.3$~arcmin$^{-2}$.

\subsection{Map Projection and Smoothing}
Because the CFHTLenS fields are irregularly shaped, and because 
we ray-trace to the actual observed galaxy positions, we first divide
them into 13 squares (subfields) to match the square shape and
$\approx$12 deg$^2$ size of our simulated maps.  Fig.~\ref{fig:
  subfields} shows the convergence maps for the CFHTLenS fields, as
well as the divisions into subfields. To maximize the data usage,
three subfields are each composed of two physically separated sky
patches (the ones with rectangular shape in the figure).

\begin{figure*}
\includegraphics[scale=0.48]{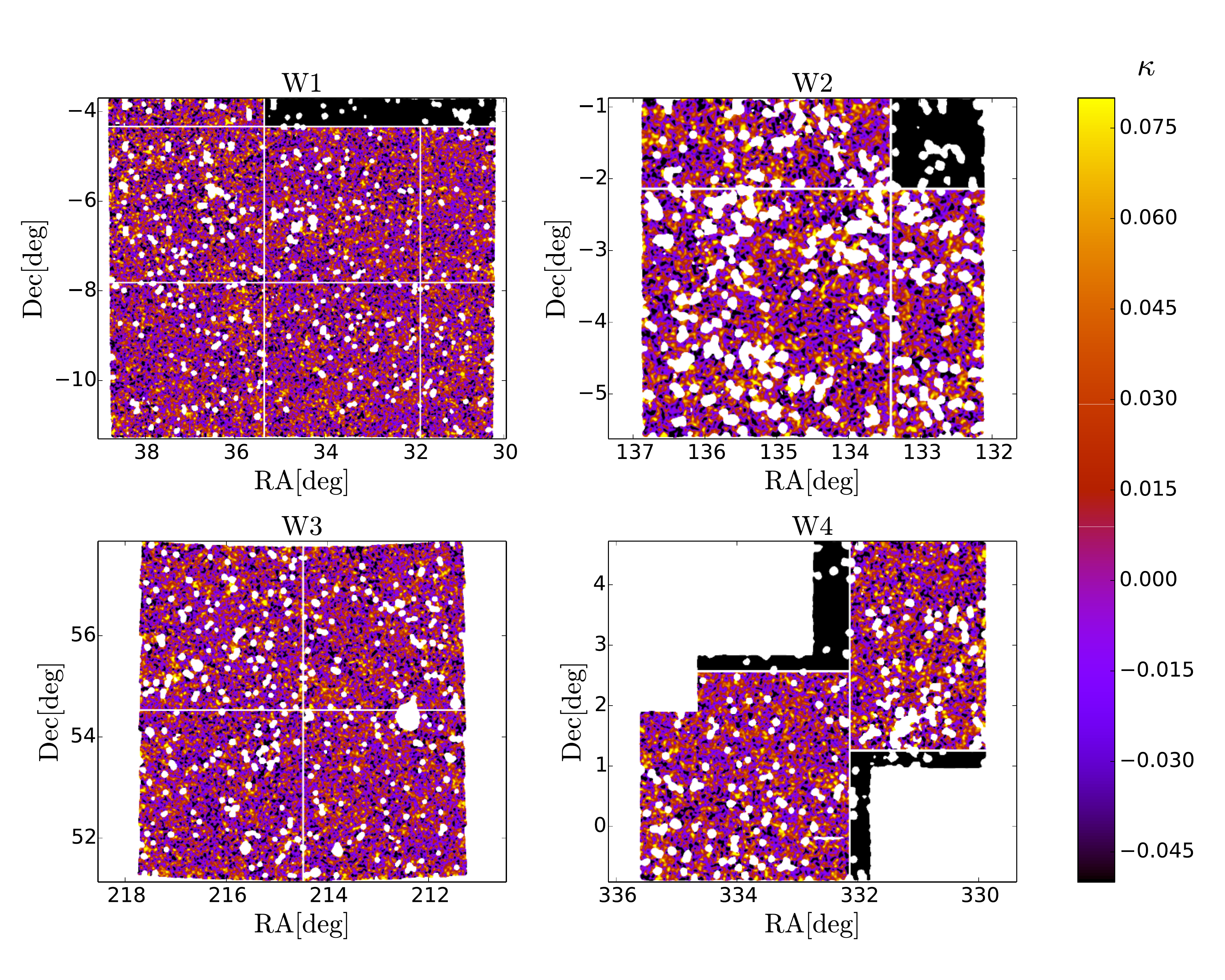}
\caption{\label{fig: subfields} Convergence maps for the four CFHTLenS
  fields. They are divided into 13 subfields of 12 deg$^2$ in size to
  match our simulation configuration. Scattered white dots are
  masks. White lines mark the edges of our simulated maps. Three
  subfields are collages of the six rectangular patches in $\rm{W1,
    W2, W4}$.  Patches in black and white are not used in our
  simulation.}
\end{figure*}

Galaxies in each subfield are then placed on a $512\times512$ pixel
grid using the flat sky (Gnomonic) projection~\cite{Coxeter1989},
\begin{eqnarray} 
x&=&\frac{\cos\phi \sin (\lambda-\lambda_0)}{\cos \eta},\\
y&=&\frac{\cos\phi_0\sin\phi-\sin\phi_0\cos\phi\cos(\lambda-\lambda_0)}{\cos \eta}
\end{eqnarray}
where $(x, y)$ is the galaxy position in radians on the grid map,
$(\lambda, \phi)$ the position in (RA, Dec), $(\lambda_0, \phi_0)$ the
center of the subfield, and $\eta$ the angular distance from the
center,
\begin{eqnarray}
\cos \eta &=& \sin\phi_0\sin\phi+\cos\phi_0\cos\phi\cos(\lambda-\lambda_0).
\end{eqnarray}

In order to reduce the noise and to perform a Fourier transform, we use a
Gaussian window function to smooth the grid map,
\begin{eqnarray}
\label{eq: e_smooth}
\overline{\eB}(\thetazB)&=&\frac{\int d^2\theta W(|\thetaB-\thetazB|)\mathcal{W}(\thetaB)\left[\eB^{\rm obs}(\thetaB)-\cB(\thetaB)\right]}
{\int\,d^2\theta W(|\thetaB-\thetazB|)\mathcal{W}(\thetaB)[1+m(\thetaB)]}\\
\label{eqn: W_G}
W(\theta)&=&\frac{1}{2\pi\theta^2_G}\exp(-\frac{\theta^2}{2\theta^2_G})\label{eq-WG},
\end{eqnarray}

where $\overline{\eB}(\thetazB)$ is the smoothed complex ellipticity
$\eB=e_1+ie_2$ at the pixel $\thetazB$. $W(\theta)$ is the Gaussian
smoothing window with scale $\theta_G$, which we choose to be 0.5,
1.0, 1.8, 3.5, 5.3, and 8.9 arcmin. $\mathcal{W}$ is the {\it lens}fit weight
for each galaxy. $c$ and $m$ are additive and multiplicative corrections,
which we include following Refs.~\cite{Heymans2012, Miller2013},
\begin{eqnarray}
\eB^{\rm obs}&=&(1+m)\eB^{\rm true} +\cB.
\end{eqnarray}

The additive correction $\cB$ is consistent with 0 for $e_1$, and
$<0.05$ for $e_2$, and $m$ is a function of signal-to-noise ($\nu_{\rm
  SN}$) and galaxy size ($r$),
\begin{eqnarray}
m(\nu_{\rm SN},r)=\frac{\beta}{\log_{10}(\nu_{\rm SN})}\exp(-\alpha r \nu_{\rm SN}),
\end{eqnarray}
with $\alpha=0.057$ and $\beta=-0.37$. This multiplicative correction
for each galaxy (denominator of eq.~\ref{eq: e_smooth}) is a fit to
the ensemble average over galaxies within the window function, because
the result can be unstable on a galaxy--by--galaxy basis when $(1+m)
\to 0$. We tested the impact of the $m$ calibration following \S~8.5
of Ref.~\cite{Miller2013}.  We sampled 100 sets of random
$(\alpha,\beta)$ values from their probability distribution provided
in Ref.~\cite{Miller2013}, and computed the variance of the power
spectrum and the peak counts among these 100 samples. Similar to the
results of the analysis in Ref.~\cite{Miller2013} for the 2PCF, we
found that this calibration impacts the power spectrum and the peak
counts at the $\lsim10^{-3}$ level, negligible comparing to the
variance between random realizations of the underlying lensing maps.

\subsection{Convergence Map Construction and Masking}\label{CFHTmass} 

The convergence ($\kappa$) and the complex shear
($\gammaB=\gamma_1+i\gamma_2$) are obtained from derivatives of the
lensing potential ($\psi$).
They are defined as
\begin{eqnarray}
\kappa(\thetaB)&=&\frac{1}{2}\nabla^2\psi(\thetaB),\\
\gamma_1(\thetaB)&=&\frac{1}{2}(\psi,_{11}-\psi,_{22}),\\
\gamma_2(\thetaB)&=&\psi,_{12},
\end{eqnarray}
where indices separated by a comma denote partial derivatives with
respect to two orthogonal components $\theta_1$ and $\theta_2$ of
$\thetaB$.  We can reconstruct the convergence map from shear
measurements \cite{Kaiser1993} using,
\begin{eqnarray}
\label{eq: KSI}
\hat{\kappa}(\ellB) = \left(\frac{\ell_1^2-\ell_2^2}{\ell_1^2+\ell_2^2}\right)\hat{\gamma}_1(\ellB)
+ 2\left(\frac{\ell_1\ell_2}{\ell_1^2+\ell_2^2}\right)\hat{\gamma}_2(\ellB)
\end{eqnarray}
where $\hat{\kappa}$, $\hat{\gamma}$ are the convergence and the shear
in Fourier space, and $\ellB$ is the wave vector with components
$\ell_1, \ell_2$. Note that ellipticity is used as a measure of the
shear, using the weak lensing approximation ($\left<\eB\right>=\gammaB$; see
below).

The data contain unusable regions (due to bright stars and bad pixels).
These regions and sky patches with low galaxy number density can
induce large errors and noise (e.g. \cite{VanderPlas2012, LiuX2014, Bard2014}). 
Hence we mask them out (shown as the
scattered white dots in Fig.~\ref{fig: subfields}). 
By masking out low density regions, we assume there is no correlation between 
the lensing signal and the galaxy number density, i.e. neglecting the magnification bias.
Ref.~\cite{Liu2014} found that the magnification bias has negligible impact 
on cosmological
parameters for surveys with $<1000$ deg$^2$ coverage.
To generate masks,  we first create
grid maps of the same size and resolution as the convergence maps, but
with each pixel containing the number of galaxies ($n_{gal}$) falling
within that pixel. We then smooth this galaxy surface density map with
the same Gaussian window function as before (Eq.~\ref{eqn: W_G}).
Finally, we remove regions where $n_{gal} < 5$~arcmin$^{-2}$ (see
Ref.~\cite{ShirasakiYoshida2014}). In order to perform a Fourier transform 
on the resulting maps, we set all pixels within the masked regions to zero. 
This can introduce noise at small scales, and we limit our final analysis to 
scales $\ell<7,000$. We also apply the same procedure on the simulated maps.

\subsection{Power Spectrum and Peak Counts}\label{PSandPK}

The power spectrum is the Fourier counterpart of the two-point
correlation function. We first Fourier transform the convergence map
(with 0.5 arcmin smoothing scale), and then average over all spherical
harmonics ($\ell=|\ellB|$) to obtain the power spectrum, with 40
equally spaced log bins in the range $370 < \ell <25,000$.

Peak counting is done by scanning through the pixels on the
convergence map, and identifying local maxima (pixels with a
higher value of $\kappa$ than its surrounding 8 pixels). We then
record the number of peaks as a function of their central $\kappa$
value. In our analysis, we use peaks with $ -0.04 < \kappa < 0.12$ 
and test various smoothing scales.

The final power spectrum is averaged over the 13 subfields, weighted
by the number of galaxies in each subfield. The final peak counts is
the sum over 13 subfields.

\section{The emulator}\label{sec: emulator}

The construction of the emulator consists of three steps. First, we
sample 91 points using the latin hypercube method 
in the three-dimensional (3D) parameter
space within the broad ranges $0<\Omega_m<1$, $-3<w<0$, and
$0.1<\sigma_8<1.5$. For each sampled point, we run an N-body
simulation and perform ray-tracing to create shear maps that are
directly comparable to the CFHTLenS data. Second, we create
convergence maps, measure the power spectra and peak counts, and
interpolate between the 91 simulated grid points to make predictions
for arbitrary cosmological models within the simulated range. Finally,
we compute the parameter likelihood in the 3D space
($\Omega_m,w,\sigma_8$) to find the best fit values and marginalized
confidence contours, using the CFHTLenS observations.

\subsection{N-body simulation and ray-tracing} 

\begin{figure}
\begin{center}
\vspace{1 in}
\includegraphics[scale=0.42]{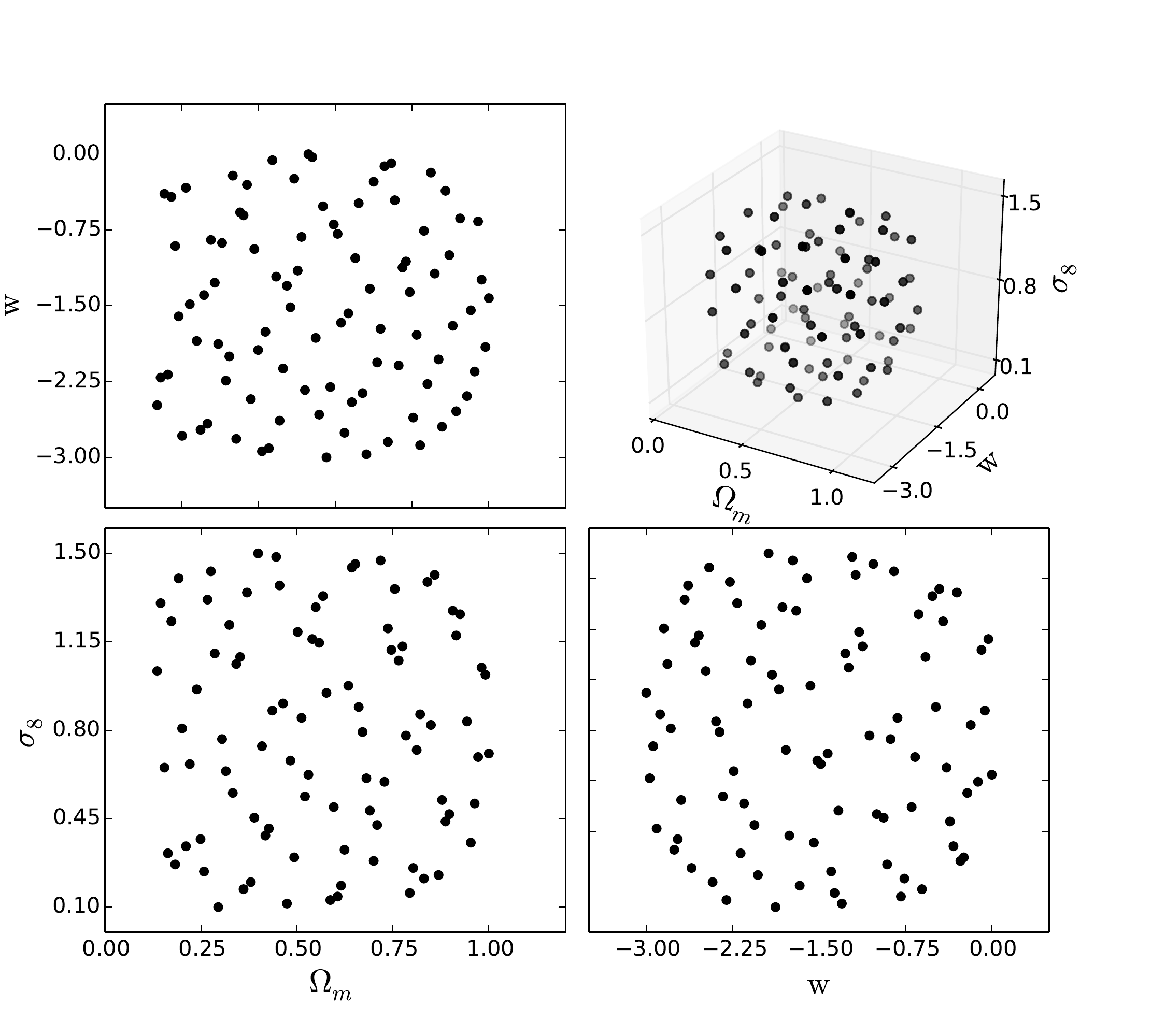}
\caption{\label{fig: cosmoparams} Visual representation of the
  cosmological parameters in the 91 models used in our simulations,
  and listed in Table~\ref{tab: CosmoParsm}.}
\end{center}
\end{figure}

We first pick 91 sampling points that are spread out in the 3D space
as evenly as possible, but not overlap when projected on 2D or 1D
space. To do this, we use the latin hypercube sampling method
following Ref.~\cite{Heitmann2009II}. A list of parameters residing on
a diagonal line is first generated, and then randomly shuffled on each
dimension. For a random pair of points and a random parameter, we swap
their values. The last step was repeated until we reached convergence
in average distance between the points ($10^5$ iterations). The
resulting parameter values are listed in Table~\ref{tab: CosmoParsm}
and shown visually in Figure~\ref{fig: cosmoparams}.

We then run one N-body simulation at each sampling point, using a
modified version of the Gadget-2
code\footnote{\url{http://www.mpa-garching.mpg.de/gadget/}}.  Except
for the values of the three cosmological parameters, the parameters
and setup of these N-body simulations are the same as used in our
earlier work \cite{Kratochvil2010, Yang2011, Kratochvil2012, Yang2013,
  Bard2013, Petri2013, Liu2014}. We refer readers to these papers for
more detailed information. The simulations have a box size of $240
h^{-1}$ comoving Mpc, containing $512^3$ dark matter particles. This
corresponds to a mass resolution of $7.4\times10^9h^{-1}M_\odot$. We
set the Hubble constant $h=0.72$, baryon density
$\Omega_{b}h^2=0.0227$, and the spectral index $n_s= 0.96$. We compute
the initial (linear) total matter power spectrum with the
Einstein-Boltzmann code CAMB\footnote{\url{http://camb.info/}}
\cite{Lewis2000} at $z=0$ and scale it back to initial redshift
$z=100$. The power spectrum is then fed into N-GenIC, the initial
condition generator associated with Gadget-2.  
Snapshot cubes are recorded at redshifts corresponding to 
every $\sim$ 80 Mpc (comoving).

To create mock shear maps, we next perform ray-tracing.  We divide
each 3D box into three parallel pieces and project each slice onto a
2D plane perpendicular to the observer's line of sight, using the
triangular shaped cloud scheme \cite{Hockney1988}. We then convert the
surface density to the gravitational potential at each plane using
Poisson's equation. At each position of the 4.2 million observed CFHTLenS galaxies, 
we follow a light ray from $z=0$, traveling backward
through the projection planes to the redshift of the galaxy,
$z_{gal}$.  For simplicity, we chose $z_{gal}$ to be the peak of the
photometric redshift probability distribution function (PDF). Using
the most probable redshift, instead of the full PDF, can potentially
induce biases as the former does not follow the stacked posterior
probabilities when fainter galaxies are included (see Fig.~10 in
Ref.~\cite{Hildebrandt2012}).  We test this effect by ray tracing to redshifts randomly drawn from the PDF of individual galaxies for one cosmology, and found the deviation of cosmological parameters to be negligible. 
Ref.~\cite{ShirasakiYoshida2014} also found the bias
caused by using the most probable photometric redshift to be small
($\Delta w_0\approx0.1$), but important for future, larger surveys.

The deflection angle, convergence, and shear are calculated at each
plane for each light ray.  Between the planes, the light rays travel
in straight lines. Finally, for each cosmological model, we create
1,000 realizations (including $\kappa$ and $\gammaB$ for each galaxy)
by randomly rotating/shifting the simulation data cubes.

In total, we created 1,183,000 mock catalogues (91 models $\times$ 13 subfields per model $\times$ 1,000 realizations per subfield).

\begin{table}
\begin{tabular}{l|c|c|c} 
\hline
 & $\Omega_m$ & w &  $\sigma_8$ \\
\hline
1 &0.624 &-2.757 &0.327\\
2 &0.849 &-0.183 &0.821\\
3 &0.136 &-2.484 &1.034\\
4 &0.295 &-1.878 &0.1\\
5 &0.418 &-1.758 &0.383\\
6 &0.615 &-1.668 &0.185\\
7 &0.558 &-2.577 &1.146\\
8 &0.915 &-2.544 &1.175\\
9 &0.7 &-0.273 &0.283\\
10 &0.446 &-1.212 &1.486\\
11 &0.991 &-1.908 &1.02\\
12 &0.155 &-0.393 &0.652\\
13 &0.145 &-2.211 &1.303\\
14 &0.981 &-1.242 &1.048\\
15 &0.409 &-2.94 &0.737\\
16 &0.436 &-0.06 &0.878\\
17 &0.183 &-0.909 &0.269\\
18 &0.502 &-1.152 &1.189\\
19 &0.38 &-2.424 &0.199\\
20 &0.887 &-0.363 &0.439\\
21 &0.276 &-0.849 &1.429\\
22 &0.718 &-1.728 &1.472\\
23 &0.755 &-0.456 &1.359\\
24 &0.831 &-0.759 &0.213\\
25 &0.455 &-2.637 &1.373\\
26 &0.671 &-2.364 &0.793\\
27 &0.765 &-2.091 &1.076\\
28 &0.493 &-0.243 &0.297\\
29 &0.483 &-1.515 &0.68\\
30 &0.474 &-1.302 &0.114\\
31 &0.84 &-2.274 &1.387\\
32 &0.963 &-2.151 &0.51\\
33 &0.258 &-1.395 &0.241\\
34 &0.972 &-0.666 &0.694\\
35 &0.943 &-2.394 &0.835\\
36 &0.643 &-2.454 &1.444\\
37 &0.821 &-2.88 &0.863\\
38 &0.775 &-1.122 &1.132\\
39 &0.54 &-0.03 &1.161\\
40 &0.352 &-0.576 &1.09\\
41 &0.333 &-0.213 &0.552\\
42 &0.897 &-0.999 &0.468\\
43 &0.221 &-1.485 &0.666\\
44 &0.953 &-1.545 &0.355\\
45 &0.315 &-2.241 &0.638\\
\hline
\end{tabular}
\quad
\begin{tabular}{c|c|c|c} 
\hline
46 &0.361 &-0.606 &0.171\\
47 &0.389 &-0.939 &0.454\\
48 &0.634 &-1.575 &0.976\\
49 &0.305 &-0.879 &0.765\\
50 &0.211 &-0.333 &0.341\\
51 &0.812 &-1.788 &0.722\\
52 &0.661 &-0.486 &0.892\\
53 &0.681 &-2.97 &0.61\\
54 &0.746 &-0.09 &1.118\\
55 &0.464 &-2.121 &0.906\\
56 &0.568 &-0.516 &1.331\\
57 &0.737 &-2.847 &1.203\\
58 &0.427 &-2.91 &0.411\\
59 &0.249 &-2.727 &0.369\\
60 &0.652 &-1.029 &1.458\\
61 &0.794 &-1.365 &0.156\\
62 &0.925 &-0.636 &1.259\\
63 &0.164 &-2.181 &0.313\\
64 &0.267 &-2.667 &1.317\\
65 &0.192 &-1.605 &1.401\\
66 &0.324 &-2.001 &1.217\\
67 &0.577 &-3.0 &0.948\\
68 &0.596 &-0.696 &0.496\\
69 &0.728 &-0.12 &0.596\\
70 &0.173 &-0.423 &1.231\\
71 &0.803 &-2.607 &0.255\\
72 &0.53 &0.0 &0.624\\
73 &0.69 &-1.332 &0.482\\
74 &0.549 &-1.818 &1.287\\
75 &0.239 &-1.848 &0.962\\
76 &0.906 &-1.698 &1.273\\
77 &0.512 &-0.819 &0.849\\
78 &0.399 &-1.938 &1.5\\
79 &0.37 &-0.303 &1.345\\
80 &0.869 &-2.031 &0.227\\
81 &0.709 &-2.061 &0.425\\
82 &0.286 &-1.272 &1.104\\
83 &0.784 &-1.062 &0.779\\
84 &0.342 &-2.817 &1.062\\
85 &1.0 &-1.425 &0.708\\
86 &0.878 &-2.697 &0.524\\
87 &0.606 &-0.789 &0.142\\
88 &0.521 &-2.334 &0.538\\
89 &0.587 &-2.304 &0.128\\
90 &0.201 &-2.787 &0.807\\
91 &0.859 &-1.182 &1.415\\
\hline
\end{tabular}
\caption[]{\label{tab: CosmoParsm} Cosmological parameters used in our
  simulations.  The universe is assumed to be spatially flat
  ($\Omega_\Lambda+\Omega_m=1$), with the Hubble constant $h=0.72$,
  baryon density $\Omega_{b}h^2=0.0227$ and spectral index $n_s=
  0.96$.}
\end{table}

\subsection{Convergence Maps}

Next, we process the simulation catalogues, mimicking as closely as
possible the procedures applied to the real CFHTLenS data. The
transformation from intrinsic to observed galaxy ellipticity
is~\cite{Seitz1997},

\begin{eqnarray}
\eB &=& \left\{
\begin{array}{c l}
\frac{\eB_{\rm int}+\gB} {1+\gB^* \eB_{\rm int}} & \quad |\gB| \leq 1\\[1.0em]
\frac{1+\gB\eB^*_{\rm int}} {\eB^*_{\rm int}+\gB^*} & \quad |\gB| > 1\\
\end{array}\right.\\
\gB&=&\frac{\gammaB}{1-\kappa},
\end{eqnarray}
where $\eB_{\rm int}$ is the galaxy's intrinsic ellipticity.
For each simulated galaxy, we assign an intrinsic ellipticity by
rotating the observed ellipticity for that galaxy by a random angle on
the sky, while conserving its magnitude $|\eB|$. $\gB = g_1+ig_2$ is
the reduced shear, and asterisk denotes complex conjugation. To be
consistent with the CFHTLenS analysis, we adopt the weak lensing limit
($|\gammaB|\ll1,\,\kappa\ll1$), hence $\gB \approx \gammaB$,
$\eB\approx\eB_{\rm int}+\gammaB$.  We estimate the bias on
cosmological parameters to be $<50\%$ of the one $\sigma$ error
for $\sigma_8$, and $<30\%$ for
$\Omega_m$, using results from Ref.~\cite{Dodelson2006} for a
CFHTLenS-like survey (with $n_{gal} = 9.3$ arcmin$^{-2}$ and sky
coverage $f_{\rm sky}=0.03$). We also add multiplicative noise by
replacing $\gammaB \to \gammaB(1+m)$.  As with CFHTLenS data, we
continue with smoothing (eq.~\ref{eq: e_smooth}), convergence map construction
(Eq.~\ref{eq: KSI}), masking (\S~\ref{CFHTmass}), and computing the
power spectrum and peak counts (\S~\ref{PSandPK}).

\subsection{Interpolation}\label{interp}

We test two methods to interpolate from the 91 measured power spectra
and peak counts to other cosmologicial models: (1) multi-dimensional
{\it Radial Basis Function} (RBF) and (2) {\it Gaussian Process}
(GP). RBF uses the average power spectrum or the peak counts (over
1,000 realizations) at each sampled point. The interpolated value is
the weighted average of all sampling points, and the weight is a
function of the distance from the interpolation point. We choose the
function to be ``multiquadric'' ($\sqrt{(p_i/\epsilon)^2 + 1}$, where
$p_i=|\pB_i-\pB_0|$ is the distance in parameter space, and $\epsilon$
is a constant chosen to be the average distance between sampling
points), as it gives us the best results among other commonly used
functions\footnote{For example, ``inverse'':~$1/\sqrt{(p/\epsilon)^2 +
    1}$, ``Gaussian'':~$\exp{[-(p/\epsilon)^2]}$, ``linear'':~$p$,
  ``cubic'':~$p^3$, and ``quintic'':~$p^5$.}. RBF interpolation is
computed using scipy\footnote{\url{http://www.scipy.org}}. 
The GP method is a technique to interpolate smooth functions on an
irregular grid, minimizing artifacts due to clustering of sampled
points in parameter space. 
GP utilizes not only the mean value at each point, but also the standard deviation among the simulated realizations.
We compute GP interpolation using the
scikit-learn package\footnote{\url{http://scikit-learn.org}}.

\begin{figure}
\begin{center}
\includegraphics[scale=0.37]{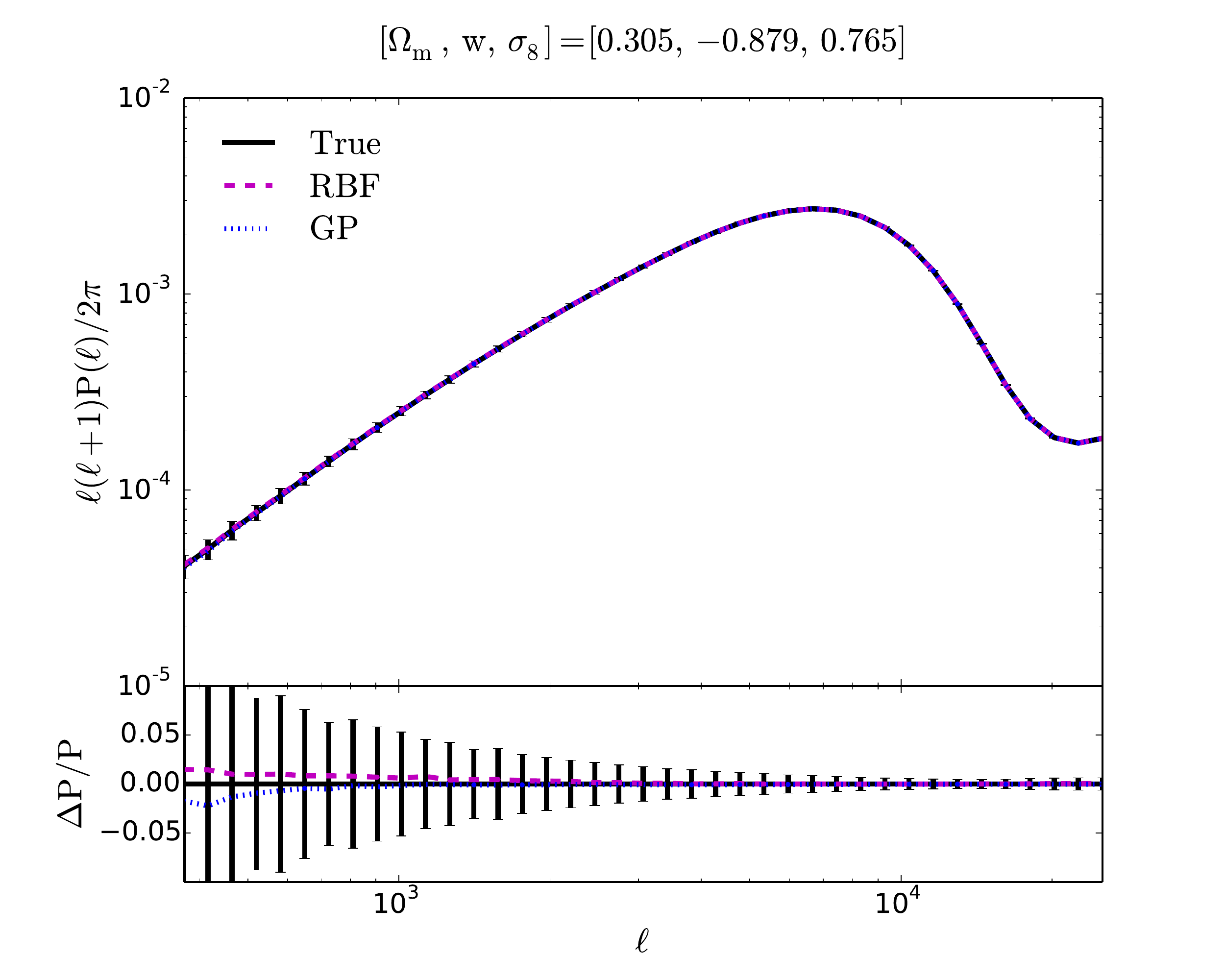}
\includegraphics[scale=0.37]{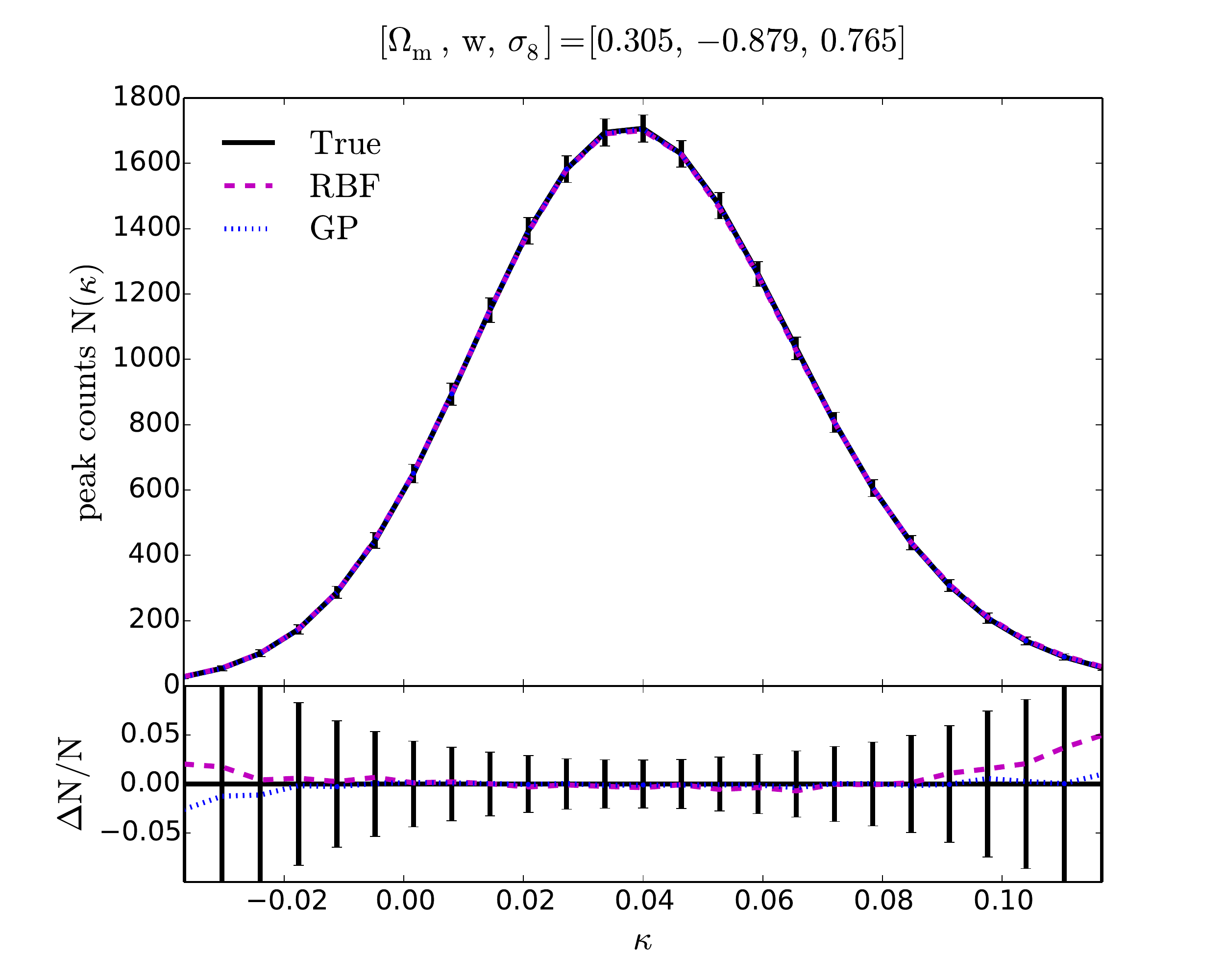}
\caption{\label{fig: sampleInterp} Examples of the interpolated power
  spectrum (upper panel) with 0.5 arcmin smoothing and peak counts
  (lower panel) with 1.0 arcmin smoothing and 25 $\kappa$ bins, using
  the two different interpolation techniques {\it Radial Basis
    Function} (RBF) and {\it Gaussian Process} (GP).  The solid curves
  show the true quantities for the given cosmological model (\#49 in
  Table~\ref{tab: CosmoParsm}), and the dashed and dotted curves show
  the interpolations based on the other 90 models.}
\end{center}
\end{figure}

Though GP uses more information than RBF, we do not find a significant
difference between the two algorithms. We test the validity of both
interpolators as follows. First, we choose one model as the test
point, and use the remaining 90 models to construct the
interpolator. We then compare the prediction at the test point to the
actual power spectrum and peak counts. This is repeated 91 times for
all models. For both power spectrum and peak counts, using either RBF
or GP, we are able to predict at $\sim1\%$ level for the power
spectrum (with only one case that is over 5\%) and at $\sim5\%$ level
for peak counts (with few cases that are slightly larger than 5\% for
high $\kappa$ peaks). Most our predictions are well within the error
bars (i.e. the variance between realizations). The interpolation
performance decreases slightly at the edges of the model parameter
space. Fig.~\ref{fig: sampleInterp} shows a typical example of the
interpolated power spectrum and peak counts, compared against the
actual values.  In our final analysis, we use RBF for faster
computation.

\subsection{Parameter Estimation}

With only three free parameters, we can directly compute the
probability distribution on a 3D parameter grid. According to Bayes's
theorem, the posterior probability of a set of parameters $\pB =
[\Omega_m, w, \sigma_8]$ for given data $\DB=[d_1, d_2, ... d_n]$
is,
\begin{eqnarray}
P(\pB | \DB) = \frac{P(\pB) P (\DB | \pB)} {P(\DB)},
\end{eqnarray}
where $P(\pB)$ is the prior, $P (\DB | \pB)$ the likelihood function
of measuring $\DB$ given $\pB$, and $P(\DB)$ the normalization.  Under
the assumption that the observables are Gaussian distributed, the
likelihood function is,
\begin{eqnarray}
P (\DB | \pB) &=& \frac{1}{(2\pi)^{n/2}|\CB|^{1/2}}  \nonumber \\
&&\times\exp\left[-0.5(\DB-\muB)\CB^{-1}(\DB-\muB)\right],
\end{eqnarray}
where $\muB$ is the prediction as described in \S~\ref{interp}, $n$ is
the number of free parameters ($=3$ in this work), and $\CB$ the
(constant) covariance matrix. We compute $\CB$ using a fiducial model
$[\Omega_m, w, \sigma_8] = [0.305, -0.879, 0.765]$, assuming
$d\CB/d\pB$ is small. The fiducial model is selected from the 91
models so that its parameters are close to the WMAP7
values~\cite{Komatsu2011}.  We use a flat prior for $\Omega_m$ in $[0,
  0.8]$, $w$ in $[-2.1, -0.3]$, and $\sigma_8$ in $[0.1, 1.4]$.
We obtain the normalization $P(\DB)$ by setting the sum of the
probability of all grid points to unity.  Within the range of our flat
priors, we compute $P(\pB | \DB)$ for $100^3$ equally spaced grid
points. To obtain 2D error contours, we marginalize over the third
parameter. The results are presented in \S~\ref{sec: results} below.

\section{Results}\label{sec: results}

\subsection{Power Spectrum}

\begin{figure*}
\begin{center}
\includegraphics[scale=0.3]{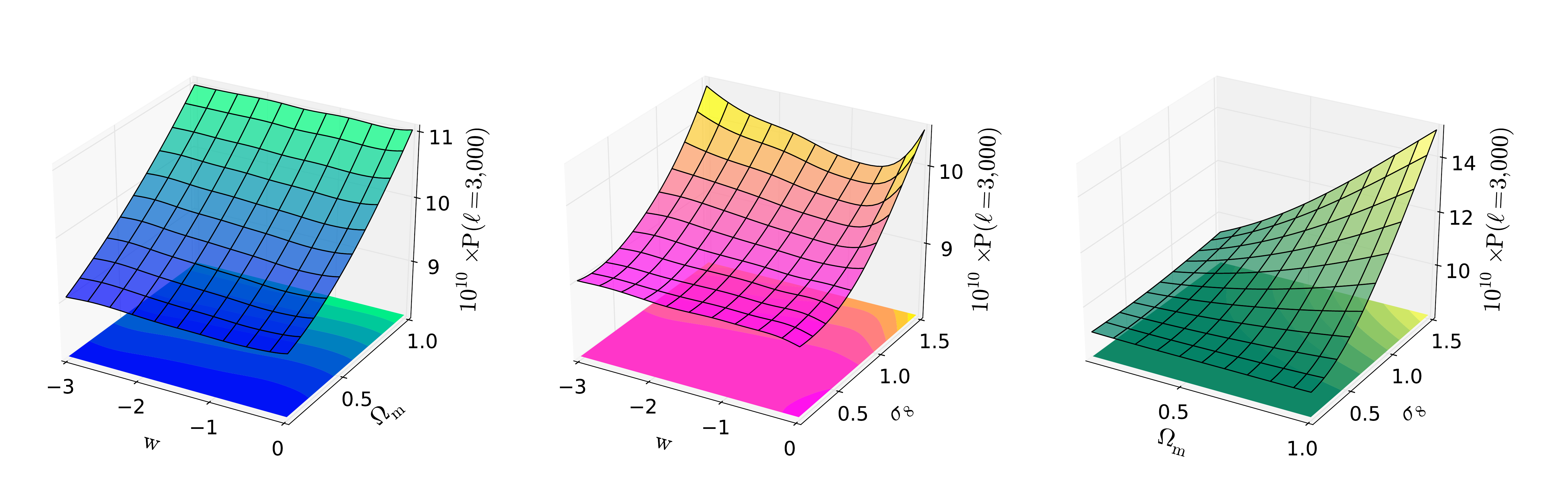}
\includegraphics[scale=0.3]{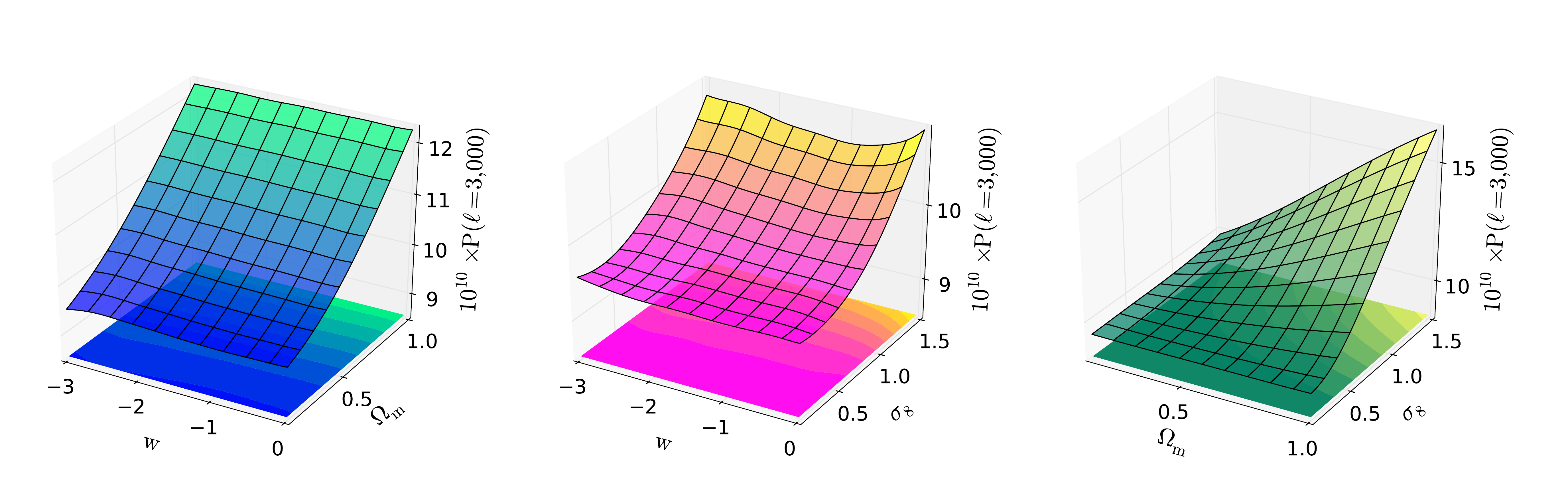}
\caption{\label{fig: PSinterpolation} Interpolated power for
  $\ell=3,000$ (7.2 arcmin) as a function of $\Omega_m$,
  $w$, and $\sigma_8$ using simulations (upper panel) and fitting
  formula from Ref.~\cite{Smith2003} (lower panel). The third
  parameter for each plane is at a fixed value ($[\Omega_m, w ,
      \sigma_8]= [0.26, -1.0, 0.8]$).}
\end{center}
\end{figure*}

We first compare our power spectrum model with theoretical
predictions. Fig.~\ref{fig: PSinterpolation} shows the interpolated
power for $\ell=3,000$ (7.2 arcmin) as a function of $\Omega_m,
w$, and $\sigma_8$. We only show the change for one particular
$\ell$, because the change is similar for all scales within our model
($370 < \ell <25,000$). The upper panel of Fig.~\ref{fig:
  PSinterpolation} is from our simulations, and the lower panel is
computed using fitting formulae from \cite{Smith2003} and the code
{\texttt
  Nicaea}\footnote{\url{http://www2.iap.fr/users/kilbinge/nicaea/}}. The
third parameter for each plane is at a fixed value ($[\Omega_m, w
    , \sigma_8]= [0.26, -1.0, 0.8]$).  Overall, simulations and
theory predictions are in good agreement, with the figure showing that
the power in the upper and lower panels is similar, and varies as a
function of cosmological parameters similarly.  For a more quantitative
test of the power spectrum, see Fig.~1 in Ref.~\cite{Yang2011}.
For cosmological constraints, we use our simulated power spectra 
directly, rather than theoretical fitting formulae.
The up-turn seen at $\ell>20,000$ is an artifact introduced by the finite 
pixel size on our maps..
However, we found no bias from this artifact, when comparing error contours 
using bins with $\ell<20,000$ and all available bins (up to $\ell=25,000$).

\begin{figure}
\begin{center}
\includegraphics[scale=0.35]{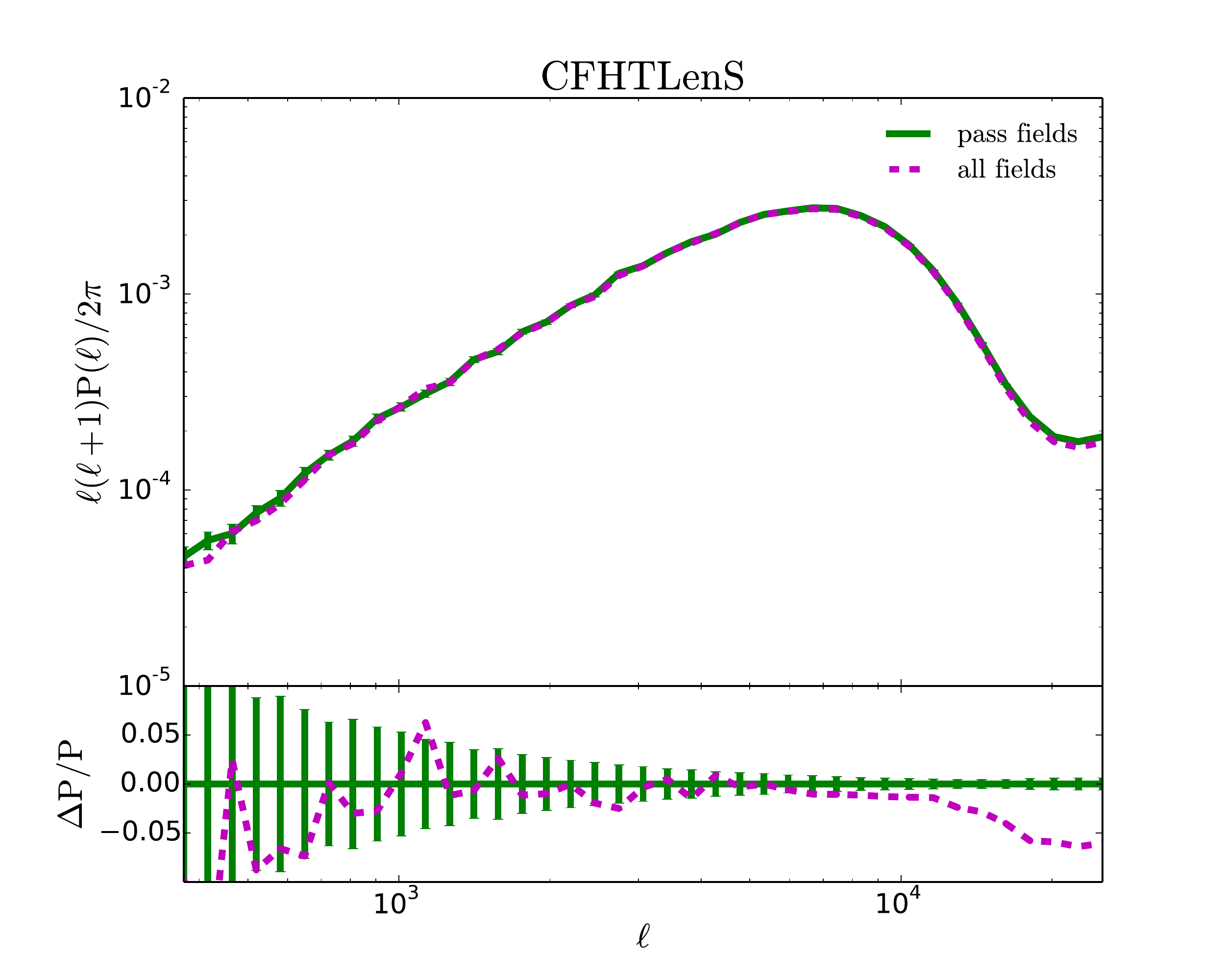}
\includegraphics[scale=0.35]{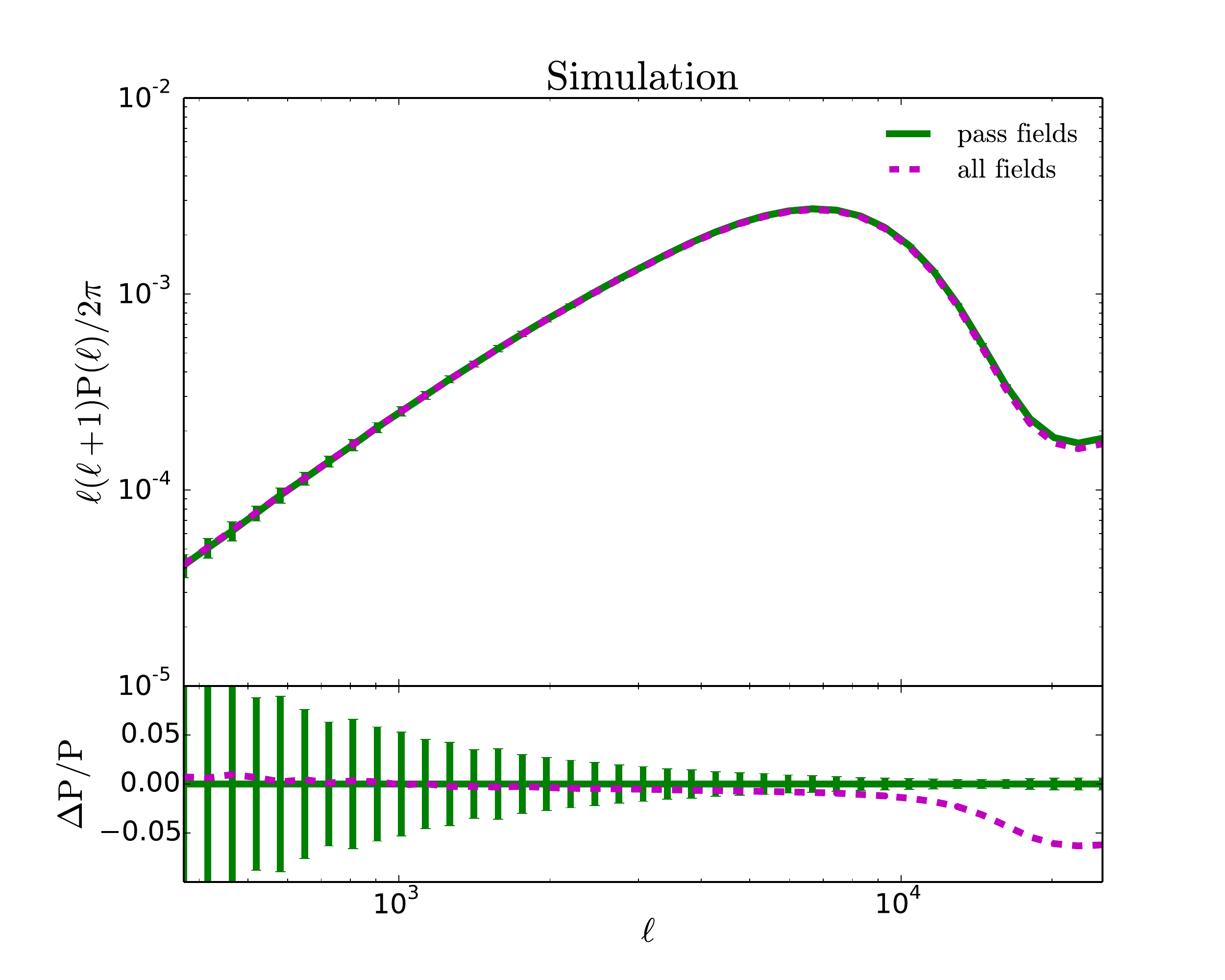}
\caption{\label{fig: ps_CFHT} Comparison of the CFHTLenS (top panel)
  and simulated (bottom panel) power spectrum measured using 75\% of
  the fields (solid line) which pass the PSF residual test and all
  fields (dashed line). The error is measured from our simulations. }
\end{center}
\end{figure}

Reference~\cite{Heymans2012} identified 25\% of the 172 individual
CFHTLenS pointings, each $\approx1$ deg$^2$ in size, with significant
PSF residuals. Including these fields can increase the systematic
error in the 2PCF, and possibly impact other statistics.  However,
because the power spectrum is a convolution of the signal and the
mask, it is susceptible to the masking pattern (due to bright stars
and bad pixels) whose characteristic scale is significantly larger
than the smoothing scale.  As each bad field removes one square degree
from the data, much larger than our $\sim$ arcmin smoothing scale,
excluding these areas can also introduce additional noise. To study
the effect due to PSF residuals and masks, we compute the power
spectrum for all fields and for the 75\% ``pass'' fields, shown in the
upper panel of Fig.~\ref{fig: ps_CFHT}. 
We find power spectra with or without
this PSF screening are consistent within errors on large scales.  On
small scales ($\ell>7,000$, or 3 arcmin), however, we find a
significant difference in power spectra with or without the corrupted
fields.  This difference is caused primarily by the particular masking
pattern, rather than field selections. This is demonstrated by
performing the same comparison using our simulated power spectra, with
the same corrupted regions either included or excluded. The result of
this comparison is shown (for the fiducial cosmology) in the lower
panel of Fig.~\ref{fig: ps_CFHT}, revealing a similarly large
discrepancy for $\ell>7,000$.  

Fig.~\ref{fig:
  ps_contour} shows the 68\% confidence level (CL) error contour in the
$\Omega_m$--$\sigma_8$ plane (marginalized over $w$) for the full
set and for the pass--only fields, and for all available $\ell$ and
for $\ell<7,000$.  We found the contours are fairly consistent among the four cases.
To be conservative, we use the 75\% pass fields only for our power spectrum analysis, 
and further limit our analysis to $\ell<7,000$.   The latter restriction eliminates small scales,
where baryonic effects can bias the shear correlation function by more than 5-10\% \cite{Semboloni2011}, and lead to a non-negligible bias on the best-constrained cosmological parameter combination
$\Sigma_8$ (defined below).
For all four contours in the figure, we are unable
to exclude the lower right corner in the $\Omega_m$--$\sigma_8$
plane. Given the strong degeneracy between $\Omega_m$ and $\sigma_8$,
it is of little meaning to quote a best fit for individual parameters; rather we
will compare the areas of the 2D contours for various probes, and obtain constraints on
a combination of the two parameters ( see below).

\begin{figure}
\begin{center}
\includegraphics[scale=0.4]{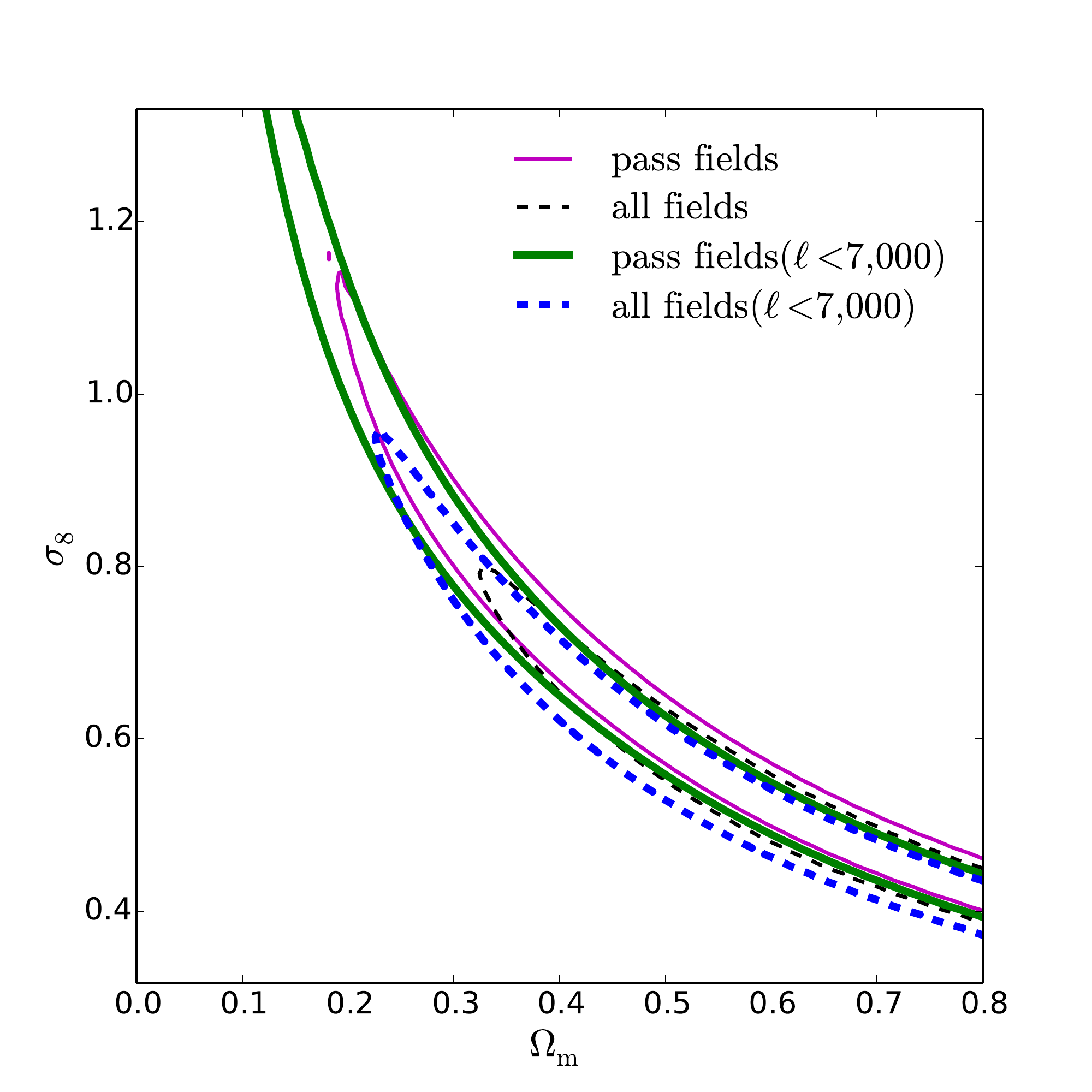}
\caption{\label{fig: ps_contour} 68\% error contours from the power
  spectrum measured using only the 75\% of the CFHTLenS fields that
  pass the PSF residual test (solid curves) and for all fields (dashed
  curves). Constraints are shown with (thick curves) and without (thin curves) 
  imposing an upper
  limit $\ell<7,000$.}
\end{center}
\end{figure}

\subsection{Peak Counts}

Interpolated peak counts from simulations as a function of
cosmological parameters are shown in Fig.~\ref{fig:
  PKinterpolation}. We present the effect for three representative
$\kappa$ values, low ($<1 \sigma_\kappa$, upper panel), medium ($1-3
\sigma_\kappa$, middle panel), and high ($>3 \sigma_\kappa$, lower
panel), where $\sigma_\kappa=0.03$ is the standard deviation in the
convergence map for 1 arcmin smoothing (with galaxy noise).  As in Fig.~\ref{fig:
  PSinterpolation}, the third parameter for each plane is at a fixed
value ($[\Omega_m, w , \sigma_8]= [0.26, -1.0, 0.8]$). Low and
high peaks behave similarly, where larger $\Omega_m$ or
$\sigma_8$ increases the number of peaks. Medium peaks behave the
opposite way. Ref.~\cite{Yang2011} investigated the origin of peaks,
and found typical high peaks are linked to one single massive halo,
while medium peaks are associated with constellations of 4--8
lower-mass, off-center halos near the line of sight. It is not
surprising to see the effect of $\Omega_m$ and $\sigma_8$ on
high peaks, as higher values increase the number of massive halos.
The opposite behavior of medium peaks is somewhat counter-intuitive,
but has been observed and explained in \cite{Yang2011}.

\begin{figure*}
\begin{center}
\includegraphics[scale=0.3]{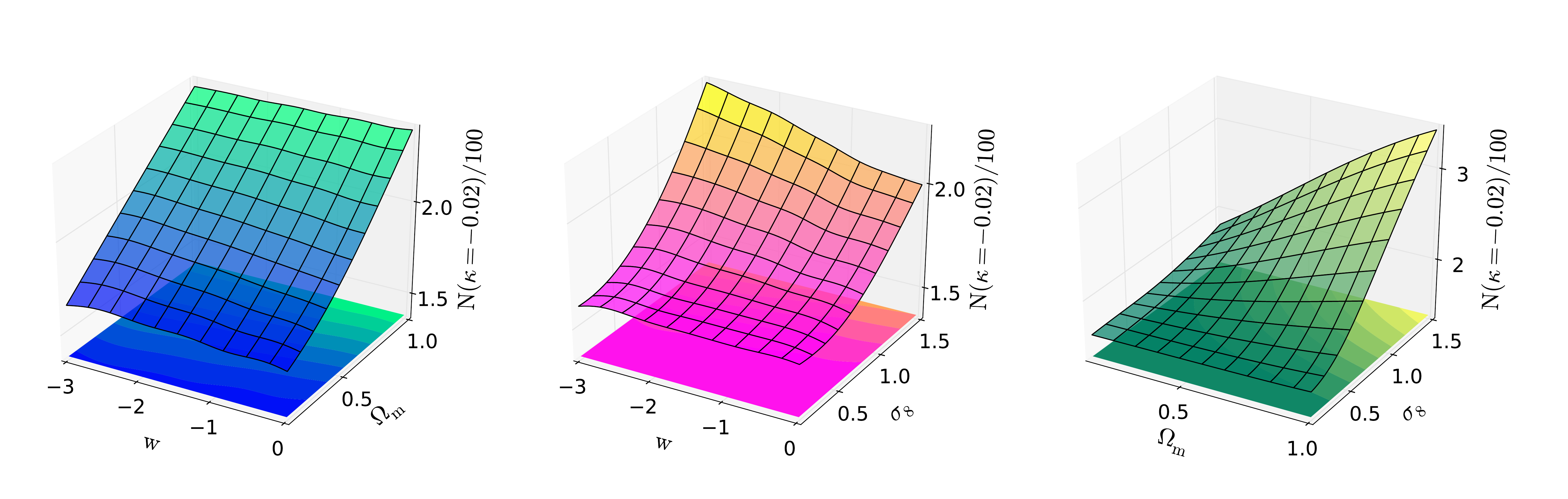}
\includegraphics[scale=0.3]{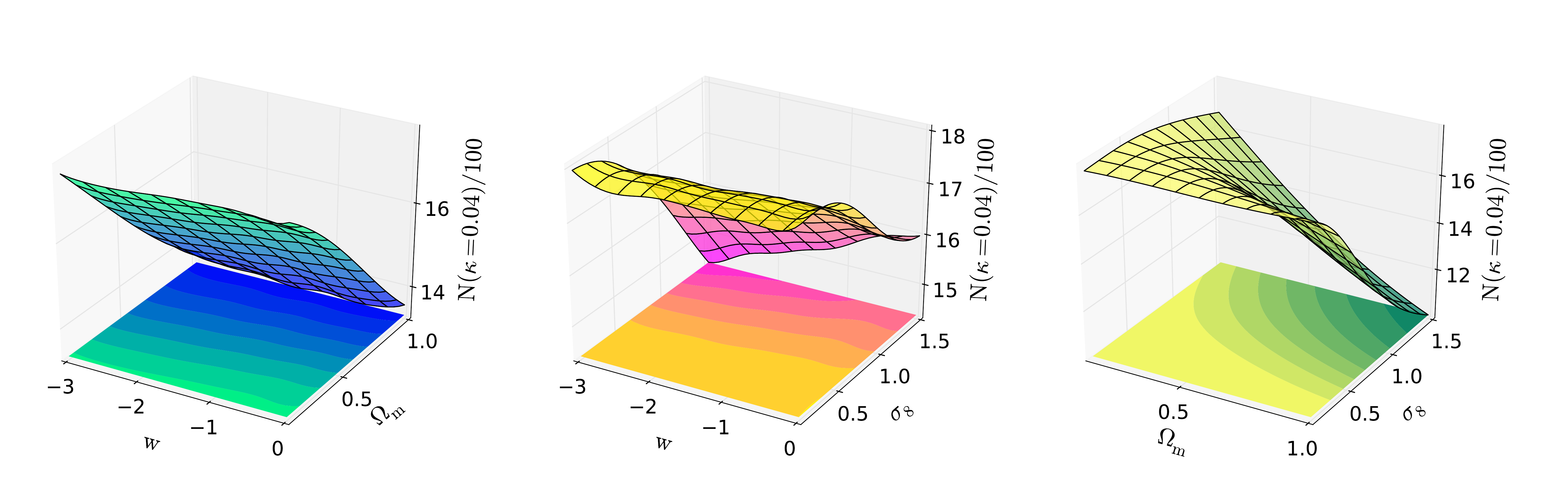}
\includegraphics[scale=0.3]{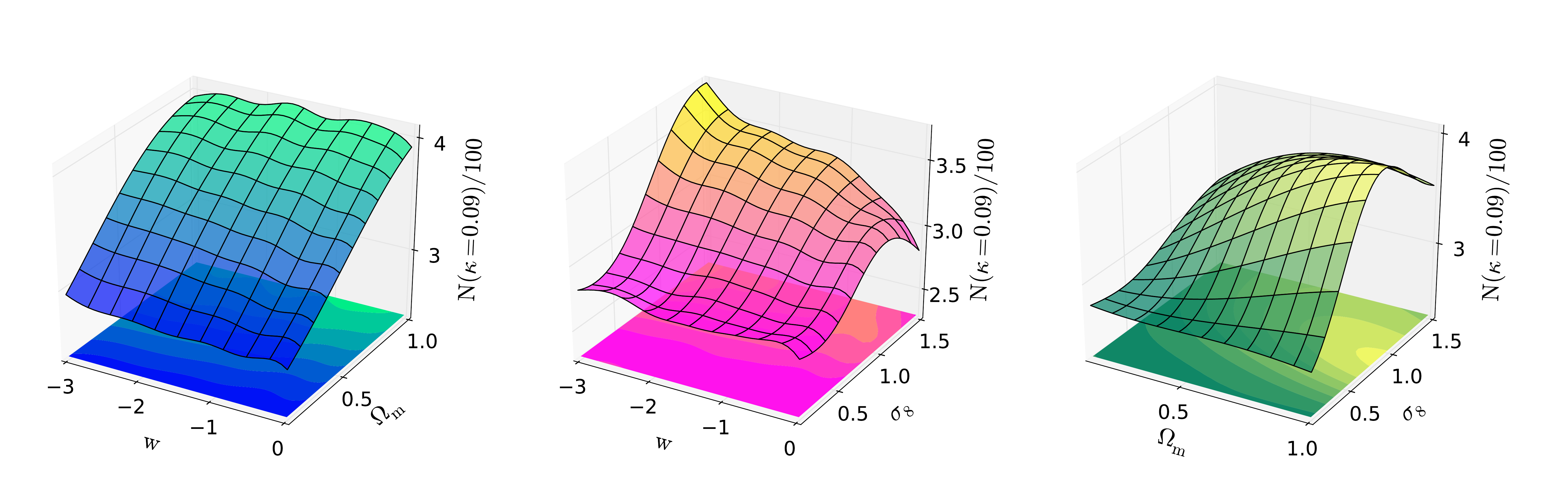}
\caption{\label{fig: PKinterpolation} Interpolated number counts for
  typical low ($<1 \sigma_\kappa$, top panel), medium ($1-3
  \sigma_\kappa$, middle panel), and high ($>3 \sigma_\kappa$, bottom
  panel) peaks, where $\sigma_{\kappa}=0.03$ is the standard deviation
  of $\kappa$ measured in our simulations.  As in Fig.~\ref{fig:
    PSinterpolation}, the third parameter in each panel is at a fixed
  value ($[\Omega_m, w , \sigma_8]= [0.26, -1.0, 0.8]$).}
\end{center}
\end{figure*}

\begin{figure}
\begin{center}
\includegraphics[scale=0.35]{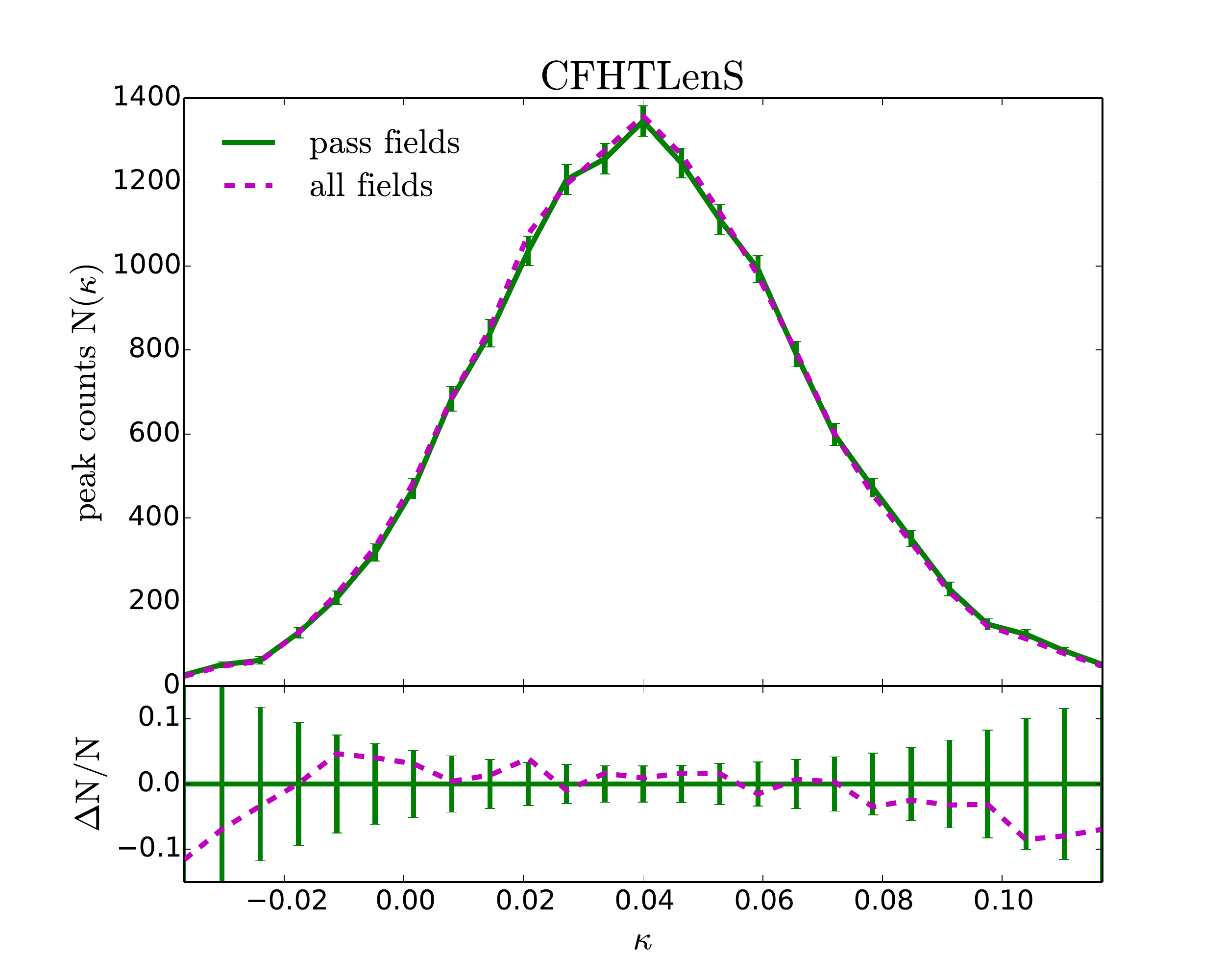}
\includegraphics[scale=0.35]{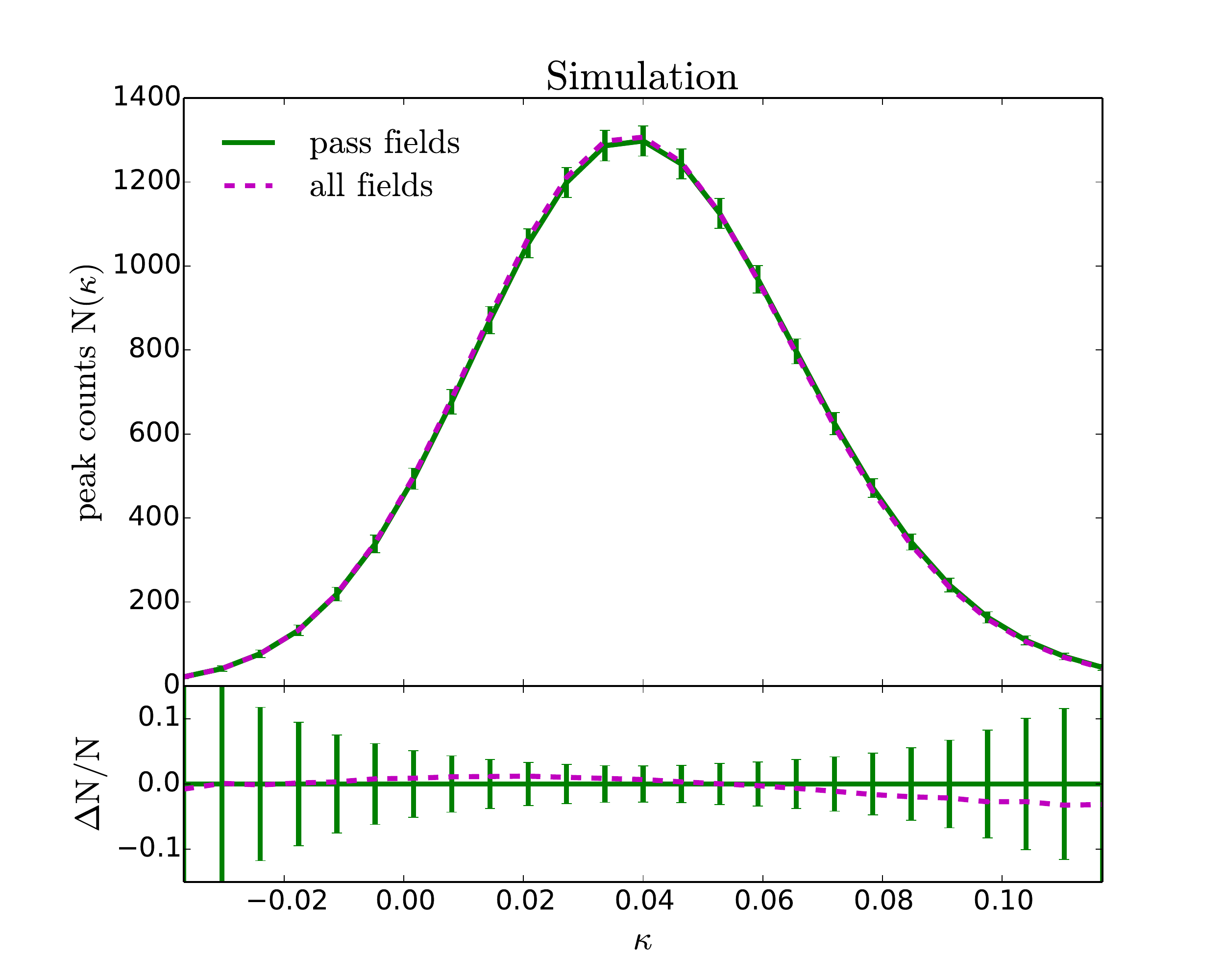}
\caption{\label{fig: pk_CFHT} Comparison of the CFHTLenS (upper panel)
  and simulated (lower panel) peak counts measured using 75\% of the
  fields (solid curves) which pass the PSF residual test and all
  fields (dashed curves). The error is measured from our simulations.}
\end{center}
\end{figure}

As peak counts are local, we expect field selections to have a smaller
impact on them, beyond modifying the total number of peaks and their
variance. This is shown to be the case in Fig.~\ref{fig: pk_CFHT},
where we compare peak counts from pass--only fields and from all
fields, and found these to be consistent for all $\kappa$ within
errors (for a fair comparison, peak counts using all fields are
multiplied by the sky ratio of pass fields to all fields, $\approx
0.75$). Therefore, unlike for the power spectrum, we choose to include
all fields for peak counts for tighter constraints.

\begin{figure}
\begin{center}
\includegraphics[scale=0.4]{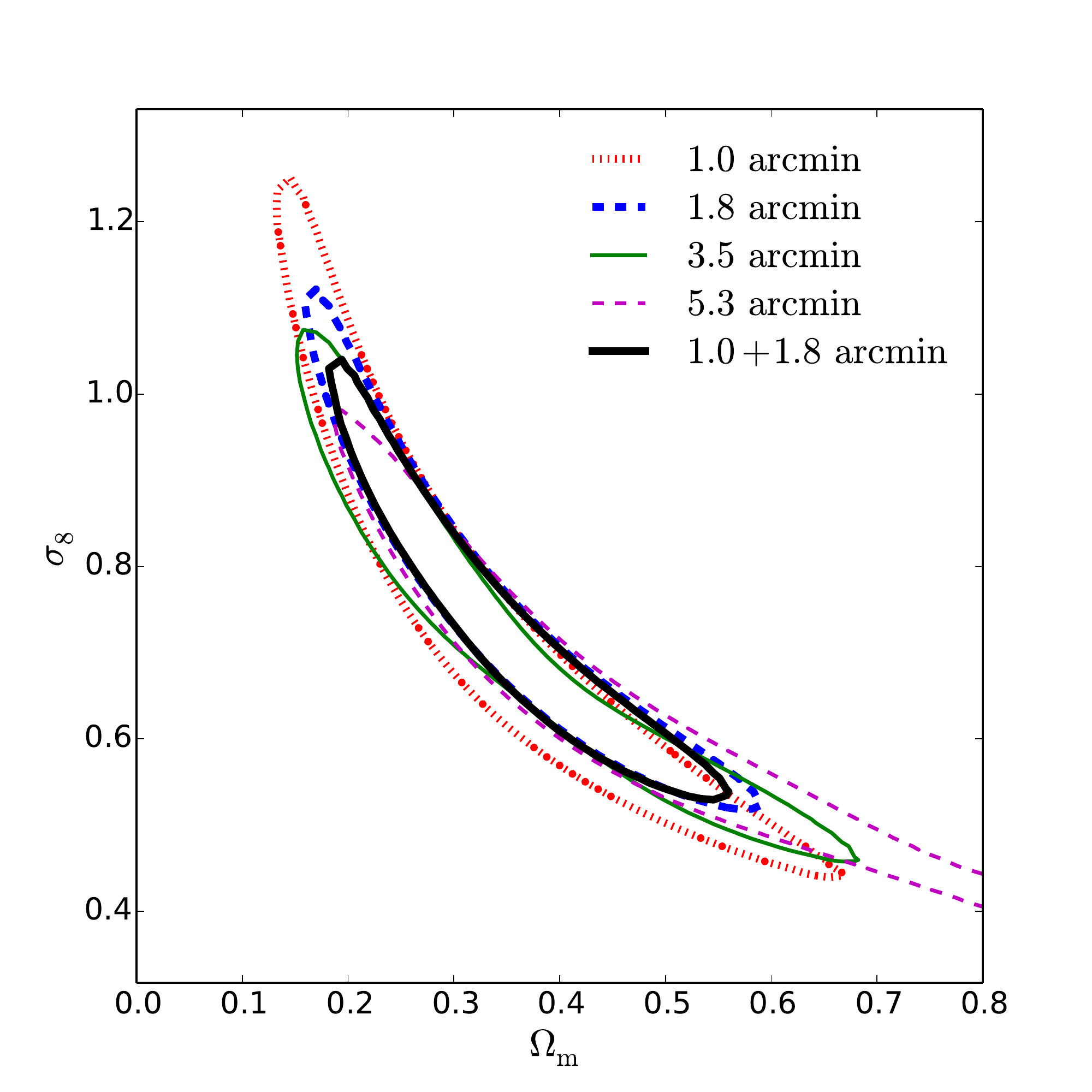}
\caption{\label{fig: peaksmoothing} 68\% error contours from peak
  counts using smoothing scales of 1.0 (dotted curve), 1.8 (thick
  dashed curve), 3.5 (thin solid curve), and 5.3 (thin dashed curve)
  arcmin, as well as from peak counts with 1.0 and 1.8 arcmin
  smoothing scales in combination (thick solid curve).}
\end{center}
\end{figure}

We test the constraints from different smoothing scales. Large
smoothing windows reduce the total number of peaks, and wash out
cosmological information, whereas small smoothing scales result in
very noisy distributions. We examine six smoothing scales (0.5, 1.0,
1.8, 3.5, 5.3, and 8.9 arcmin). The smallest (0.5 arcmin) and the
largest (8.9 arcmin) yield significantly larger errors than the other
four. We show the error contours from these four, intermediate
smoothing scales in Fig.~\ref{fig: peaksmoothing}. The 1.0 and 1.8
arcmin scales yield tighter contours than the other two, larger
scales. A combination of these two best scales (also shown in the
figure) further tighten the errors, and we therefore use it in our
final analysis. Clearly, the above is only a limited investigation of
the benefit of using multiple smoothing scales. We expect that a more
rigorous study in the future, identifying optimal filter shapes,
sizes, and combinations can help further tighten constraints from peak
counts.

\subsection{Cosmological Constraints}

From the interpolated planes for the power spectrum (Fig.~\ref{fig:
  PSinterpolation}) and peak counts (Fig.~\ref{fig: PKinterpolation}),
we see some similarity between the two statistics. They both suffer a
similar $\Omega_m$--$\sigma_8$ degeneracy, and both have a much
weaker dependence on $w$ than on $\Omega_m$ or
$\sigma_8$. However, peak counts are less impacted than the power
spectrum by field selections (due to PSF residuals) and masks, two
non-trivial systematics in CFHTLenS observations.

\begin{figure}
\begin{center}
\includegraphics[scale=0.45]{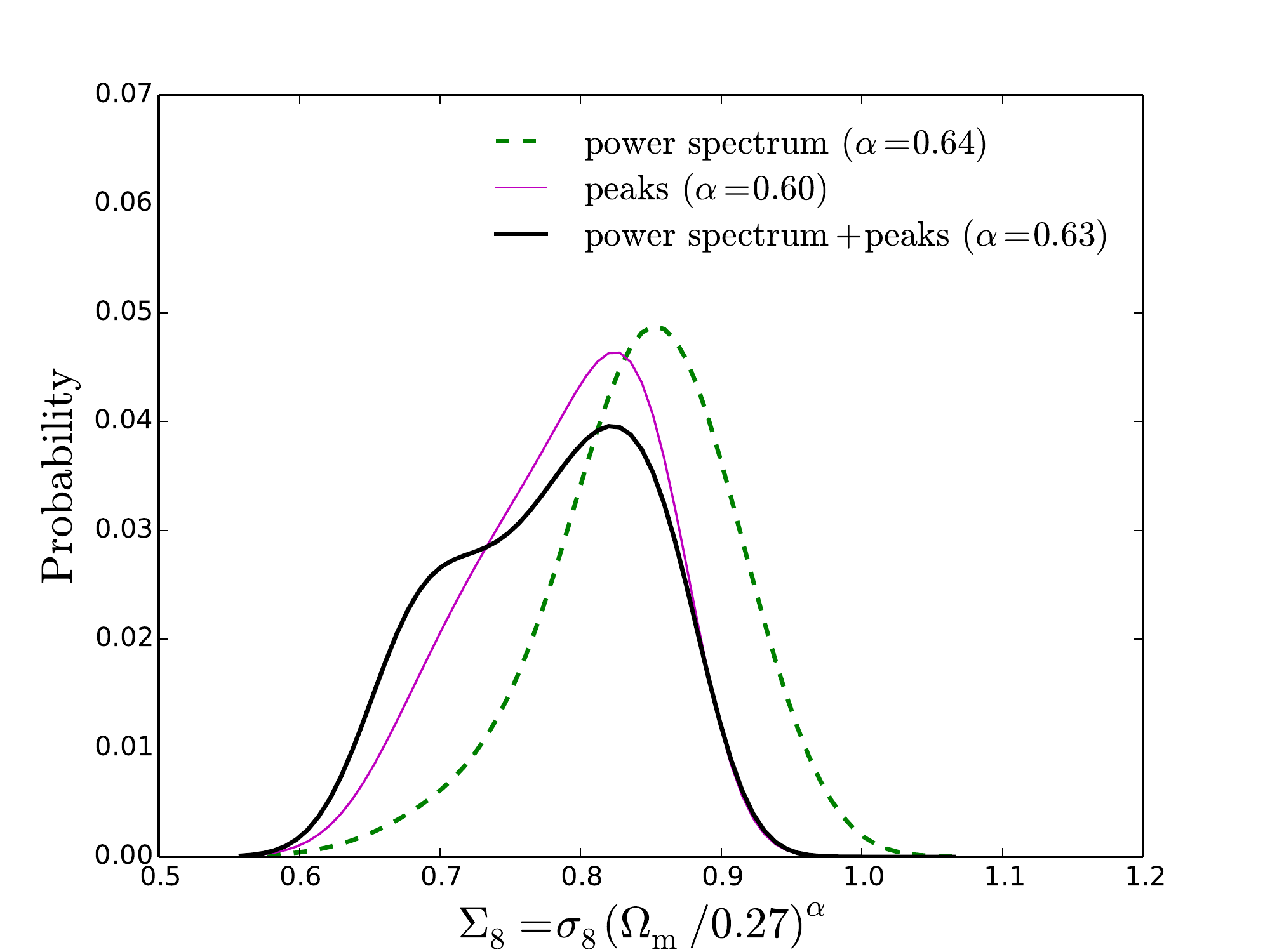}
\caption{\label{fig: SIGMA} Constraints on the parameter
  $\Sigma_8=\sigma_8(\Omega_m/0.27)^{\alpha}$, using the power
  spectrum (dashed line), peak counts (thin solid line), and their
  combination (thick solid line). The power spectrum is computed using
  the 75\% of the fields that pass the PSF residual test, and
  restrictd to $\ell<7,000$. Peak counts are computed using all
  fields, and include measurements of peaks on two smoothing scales
  (1.0 and 1.8 arcmin).}
\end{center}
\end{figure}

\begin{figure}
\begin{center}
\includegraphics[scale=0.4]{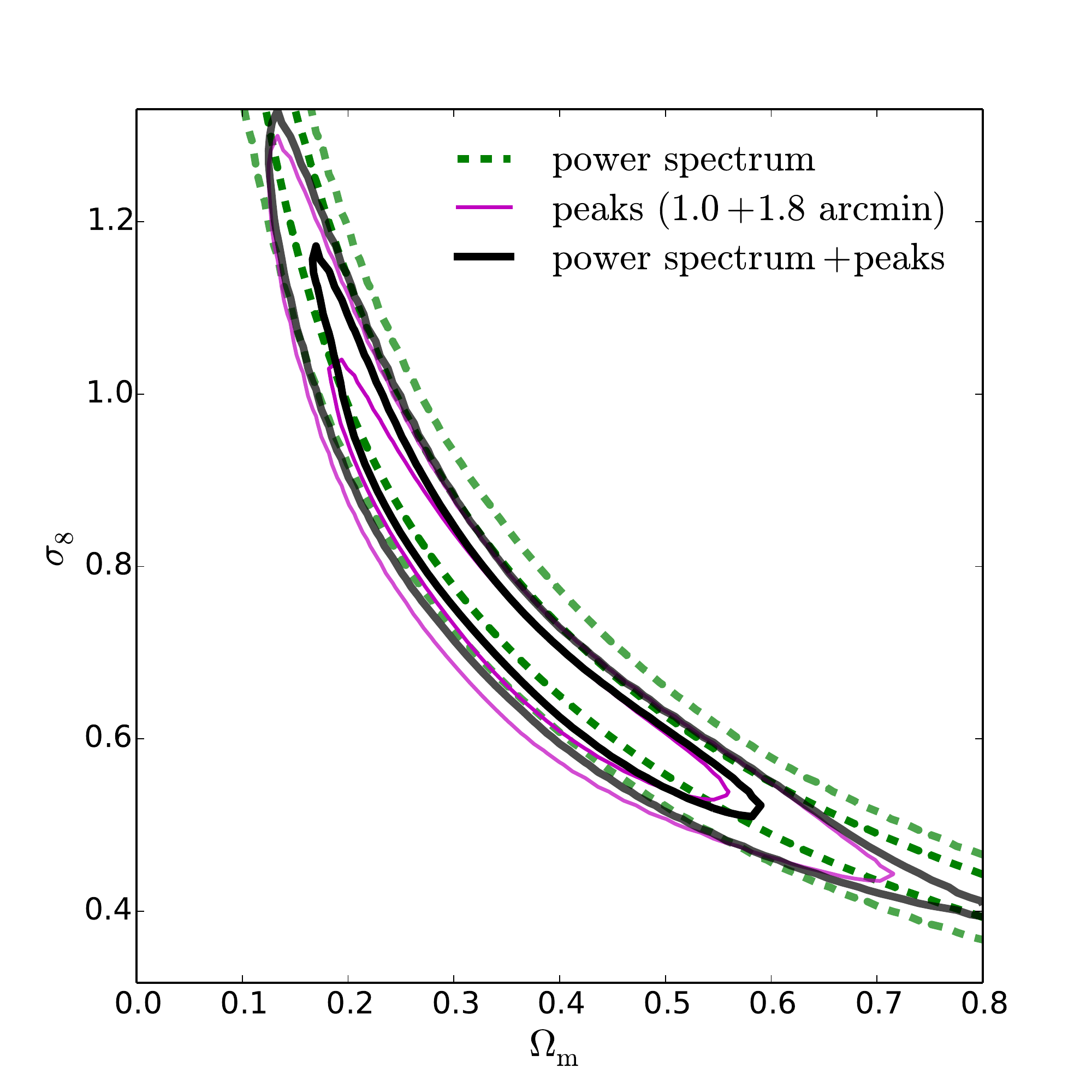}
\includegraphics[scale=0.4]{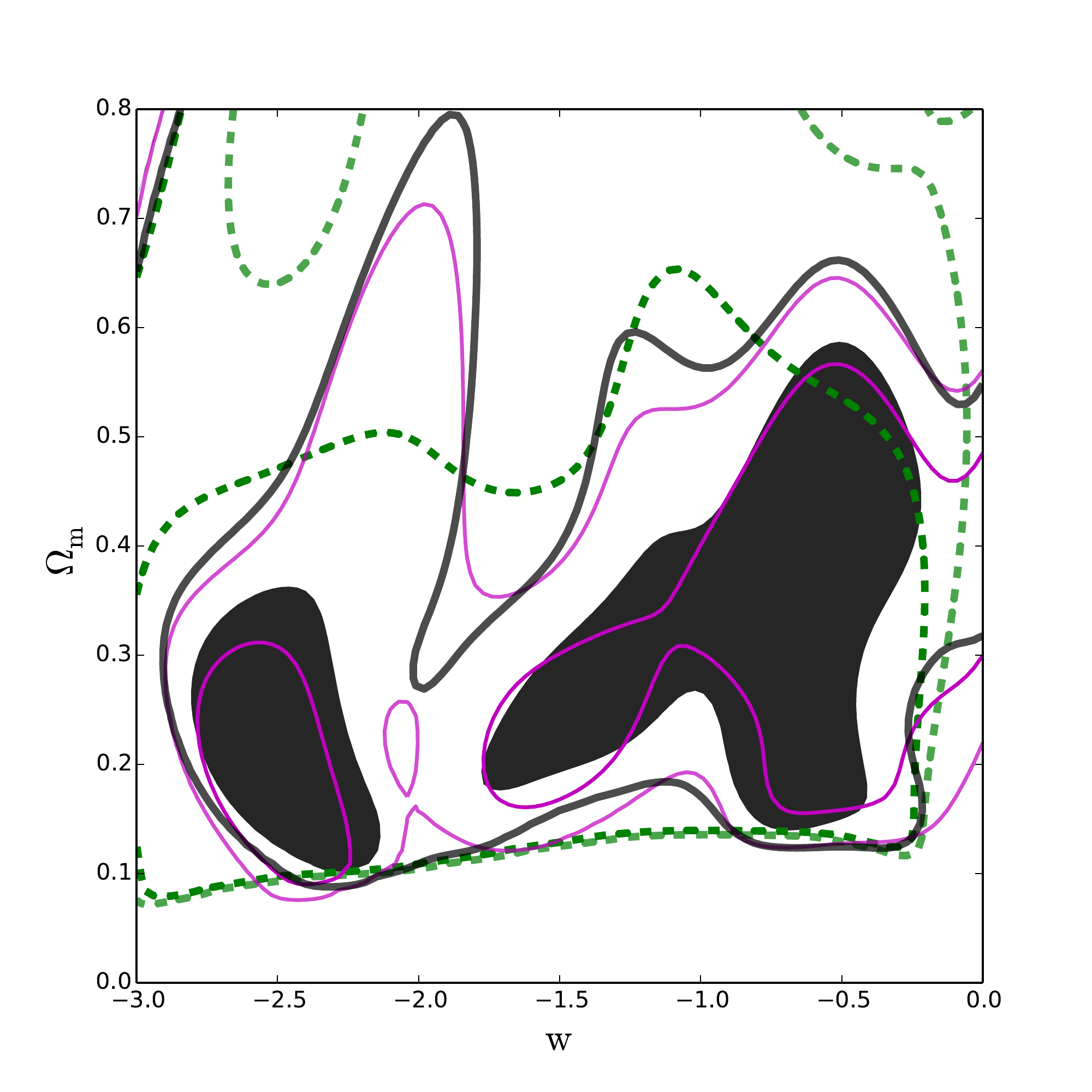}
\caption{\label{fig: pspk} 68\% (dark color) and 95\% (light color)
  error contours from the power spectrum (dashed curves), peak counts
  (thin solid curves), and their combination (thick solid curves). The
  shaded region in the bottom panel is the 68\% error contour for the
  combination. The power spectrum is computed using the 75\% of the
  fields that pass the PSF residual test, and restricted to
  $\ell<7,000$. Peak counts are computed using all fields, and include
  measurements of peaks on two smoothing scales (1.0 and 1.8 arcmin).}
\end{center}
\end{figure}

Fig.~\ref{fig: pspk} shows 68\% and 95\% confidence contours for the
power spectrum, peak counts ($1.0+1.8$ arcmin), and the combination of
both statistics.  The full covariance is taken into account when combining the 
two statistics. Table~\ref{tab: SIGMA} lists the marginalized
constraints on $\Sigma_8=\sigma_8(\Omega_m/0.27)^{\alpha}$, which
is roughly orthogonal to the $\Omega_m$--$\sigma_8$ degeneracy
direction.  We find the best fit $\alpha=0.63$ and 
$\Sigma_8=0.85\substack{+0.03 \\ -0.03}$ (with a fixed $\alpha$).
For comparison, using the 2PCF, Ref.~\cite{Kilbinger2013}
found this constraint (with best fit $\alpha=0.59$) to be $0.79\substack{+0.07 \\ -0.06}$,
 comparable
to within  $\approx 1 \sigma$ with our result (although their values have been 
marginalized over additional cosmological parameters).
Our probability distribution for $\Sigma_8$
(Fig.~\ref{fig: SIGMA}) also shows a somewhat asymmetric shape, with a
long tail to low values, when using the power spectrum, which creates
our asymmetric error bars. 

The relative area covered by each contour is listed in Table~\ref{tab: ContourSize},
normalized by the size of the 68\% contour from the power spectrum.  In both 2D parameter planes shown, the constraints from the peak counts are stronger than from the power spectrum, and largely determine the size and shape of the combined contour.
 The size of this combined
contour is a factor of $\approx 1.5-2$ smaller than from the power
spectrum alone.  One may worry that this result is unfair, as our
power spectrum analysis uses only 75\% of all fields and is restricted
to $\ell < 7, 000$, while peak counts use all fields and include
information from smoothing scales as small as 1 arcmin.  We find that
using all fields can reduce the power spectrum error contour by 83\%,
while using all available $\ell$ can reduce the contour by 90\%.  When
both of these restrictions on the power spectrum are lifted, the area
enclosed by the 68\% confidence level contour from the power spectrum
is 62\% smaller than that listed in Table~\ref{tab: ContourSize}, making the
power--spectrum--alone and the peaks--alone constraints comparable.
However, as argued above, the power spectrum result in this case may
be significantly biased by systematic errors and baryonic effects
(and, as shown by the blue curve in Fig.~\ref{fig: ps_contour}, the concordance
$\Lambda$CDM model is indeed outside the 68\% CL in this case).

\begin{figure}
\begin{center}
\includegraphics[scale=0.45]{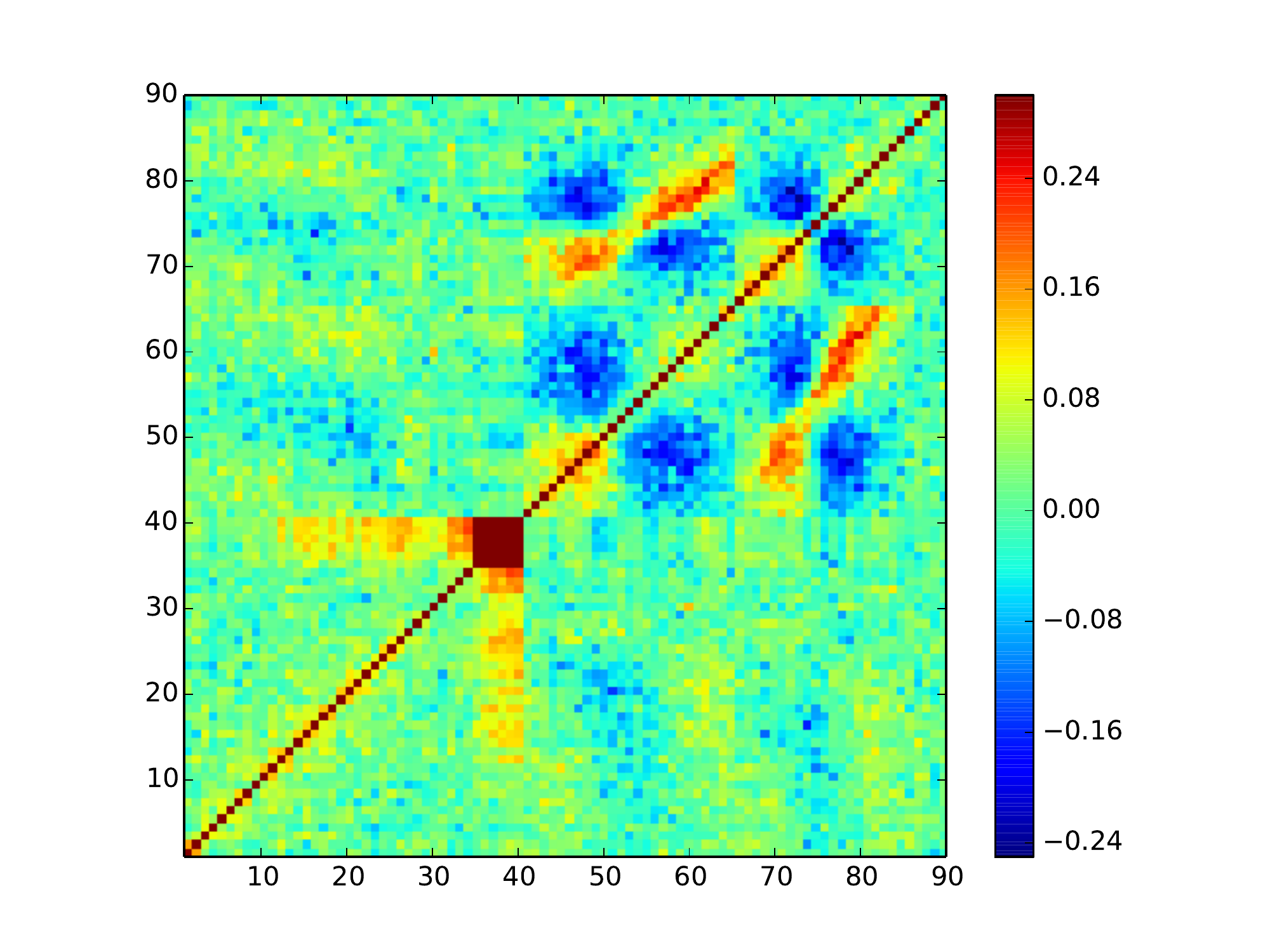}
\caption{\label{fig: corrcoeff} Correlation coefficients of the total covariance.
 Bins 1 - 40 are for the power spectrum, bin 41 - 65 are for peak counts with
 1.0 arcmin smoothing scale, and bins 66 - 90 are for peak counts with 1.8
 arcmin smoothing scale.}
\end{center}
\end{figure}

\begin{figure}
\begin{center}
\includegraphics[scale=0.4]{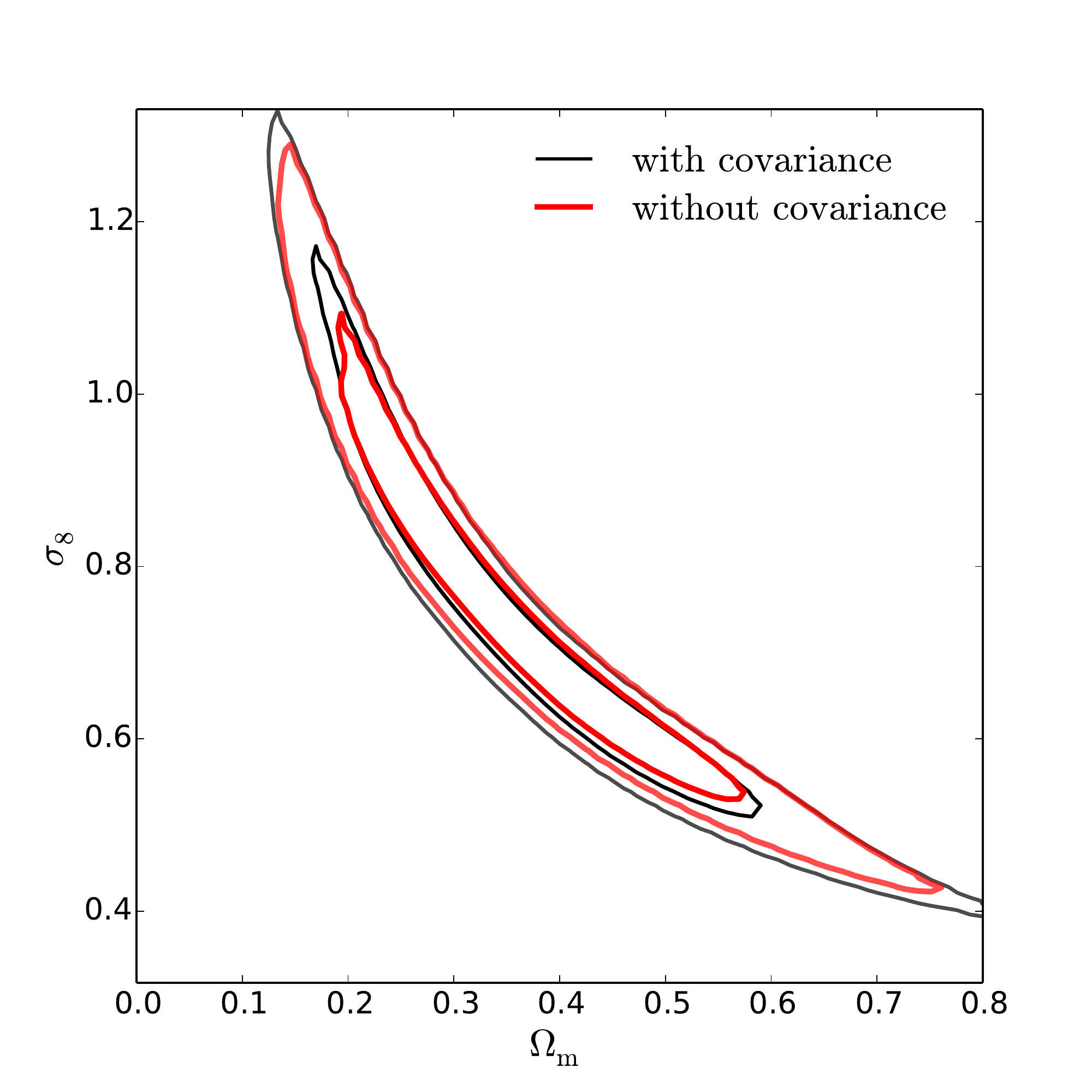}
\caption{\label{fig: pspk_cov} 68\% (dark color) and 95\% (light color)
  power spectrum + peak counts error contours 
  with (thin curves) and without (thick curves)
  the cross-covariance.  The power spectrum is computed using the 75\% of the
  fields that pass the PSF residual test, and restricted to
  $\ell<7,000$. Peak counts are computed using all fields, and include
  measurements of peaks on two smoothing scales (1.0 and 1.8 arcmin).}
\end{center}
\end{figure}

Ref.~\cite{TakadaBridle2007} examined the covariance
between cluster counts and weak lensing power spectrum and found
that including the cross-covariance leads to degradation of cosmological 
constraints by few percent
(also see Ref.~\cite{TakadaSpergel2014}). We test the importance of the 
covariance between peak counts and the power spectrum.  
Fig.~\ref{fig: corrcoeff}
shows the total covariance of the power spectrum and peak counts (1.0 + 1.8 arcmin
smoothing scales). 
Fig.~\ref{fig: pspk_cov}  shows the error contours when such cross-covariance is
included in the analysis (as done throughout our paper; black solid
curves) or ignored (dashed red curves). 
In the latter case, i.e. when the two statistics are assumed to be independent,
 the area of the 68\% CL contour is reduced by $\approx 16\%$, a somewhat
 larger change than was found for the combination of cluster counts and power 
 spectrum (although for different parameters; see Fig.12 in Ref.~\cite{TakadaBridle2007}). 

\begin{figure}
\begin{center}
\includegraphics[scale=0.35]{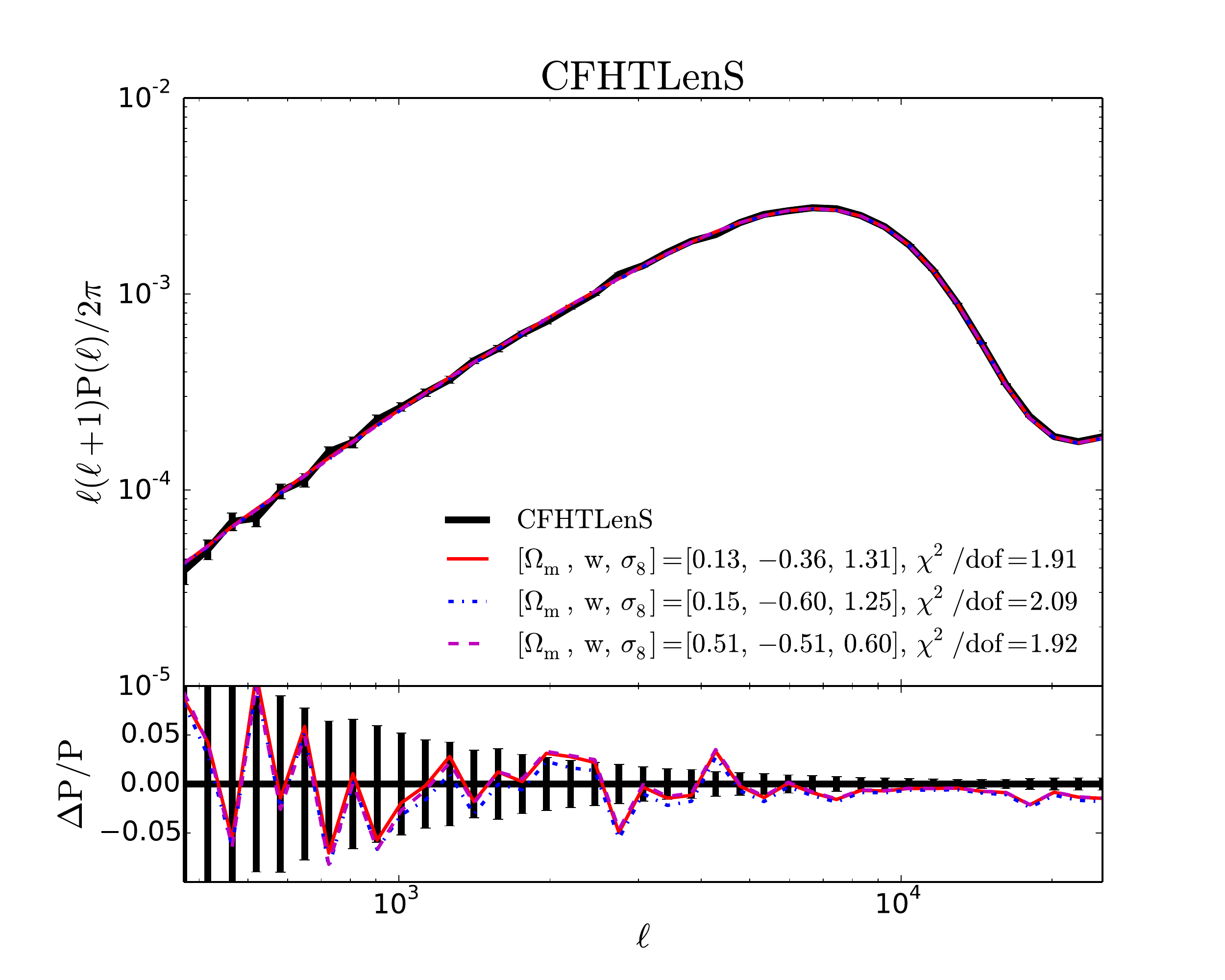}
\includegraphics[scale=0.35]{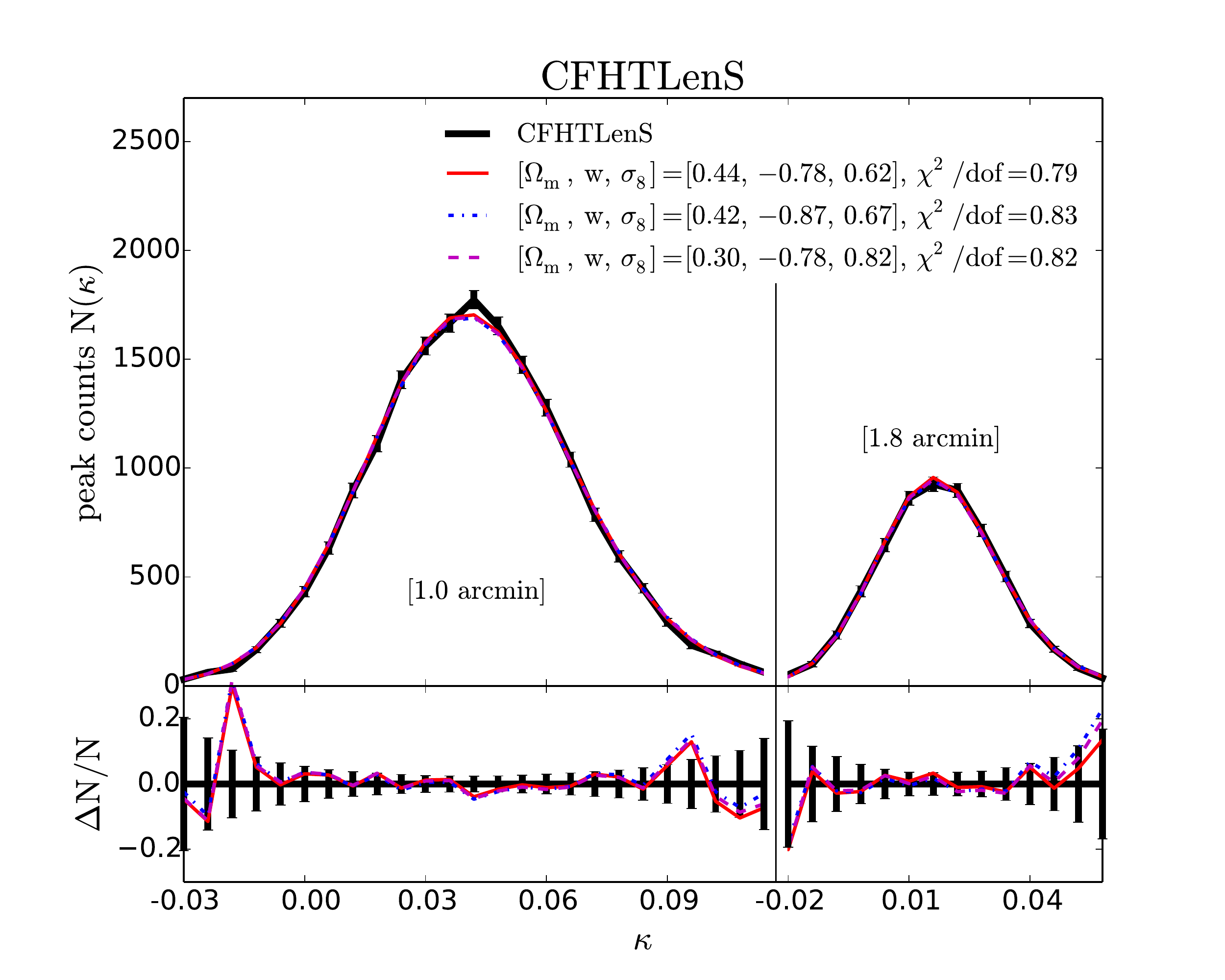}
\caption{\label{fig: pspk_fit} Fits to the CFHTLenS (thick solid
  curves) power spectrum (upper panel) and peak counts (lower
  panel). The peak counts on 1.0 and 1.8 arcmin scales are
  concatenated on the x-axis. The best fits (thin solid curves) and
  two other models (dash-dotted and dashed curves) randomly selected
  from within the 68\% error bananas, are shown for reference.}
\end{center}
\end{figure}

Finally, we show in Fig.~\ref{fig: pspk_fit} the best--fit and two
other models, randomly selected from within the 68\% error banana,
along with the CFHTLenS power spectrum and peaks. The reduced
$\chi^2\approx 2$ for the best--fit model to the power spectrum is
large, indicating
the model does not fully describe the data, and the discrimination
between the best-fit model and other models located along the ridge of
the degeneracy ``banana'' is weak. Overall, these results indicate
that there may still be significant systematic errors, even after the
problematic fields have been excluded.  The reduced $\chi^2\approx
0.8$ for the fits to the peak counts is significantly lower.

\begin{table}
\begin{tabular}{|l|c|c|} 
\hline
& $\Sigma_8$	&	$\alpha$ \\
\hline
power spectrum	&	$0.87\substack{+0.05 \\ -0.06}$	& 	0.64	\\
peak counts		&	$0.84\substack{+0.03 \\ -0.04}$	&	0.60 \\
combined			&	$0.85\substack{+0.03 \\ -0.03}$	& 	0.63 \\
\hline
\end{tabular}
\caption[]{\label{tab: SIGMA} Marginalized 68\% constraints for
  $\Sigma_8={\sigma_8(\Omega_m/0.27)^{\alpha}}$, using the power
  spectrum, peak counts, and their combination. }
\end{table}

\begin{table}
\begin{tabular}{|l|cc|cc|} 
\hline
& \multicolumn{2}{c|} { $w$--$\Omega_m$}&\multicolumn{2}{c|}{$\Omega_m$--$\sigma_8$}\\
\cline{2-5}
				&	68\%	&	95\%	&	68\%	&	95\%	\\
\cline{1-5}
power spectrum	&	1.00	&	1.74	&	1.00	&	1.99	\\
peak counts		&	0.41	&	1.01	&	0.59	&	1.51	\\
combined			&	0.42	&	1.05	&	0.61	&	1.46	\\
\hline
\end{tabular}
\caption[]{\label{tab: ContourSize} The areas of the two-dimensional
  error contours computed using the power spectrum, peak counts, and
  their combination, in two parameter planes (marginalized over the
  third parameter). The areas are normalized to the 68\% power
  spectrum contour in each case. }
\end{table}

 added plot

\section{Conclusions}\label{sec: conclusion}

In this paper, we have run 91 cosmological models, built a
CFHTLenS--specific weak lensing emulator for the power spectrum and
peak counts, and obtained constraints on $\Omega_m$, $w$,
and $\sigma_8$. Peak counts as a recently developed non-Gaussian
statistics have previously been proven in theory to have comparable
constraining power as the power spectrum. This work is the first
attempt to test this hypothesis rigorously on real data.

We have found that combining peak counts with the power spectrum 
can reduce the area of the 2D
error contour by a factor of $\approx 2$ compared to using the power spectrum
alone. Combining both statistics, we obtained
$\sigma_8(\Omega_m/0.27)^{0.63}=0.85\substack{+0.03 \\ -0.03}$. 

To conclude, peak counts can serve as a complementary
probe to the power spectrum in two important ways:

(1) As a calibration tool for systematics. Peaks 
with small ($\sim$ arcmin) smoothing scales 
suffer less (or are impacted differently by) systematics than the power spectrum. 
For CFHTLenS, we have found that the PSF residuals have little impact on peak counts,
in contrast with the bias seen with the 2PCF in Ref.~\cite{Heymans2012}.
We find that masking also has little impact on the peak counts, whereas it changes 
the power spectrum at small scales ($\ell>7,000$). The change in the power spectrum
does not impact cosmological constraints, as long as the mask is taken into consideration
in the model (e.g. Fig.~\ref{fig: ps_contour}).
Furthermore, in our previous work on
theoretical systematics due to the magnification bias \cite{Liu2014},
we also discovered that, while both the power spectrum and peak counts
are affected, the resulting directions of the biases in the cosmology
parameter space are different. Combining the two probes can mitigate
the impact from these systematics.

(2) By providing tighter constraints on cosmological parameters. The
peak counts by themselves have a similar, or even better constraining
power than the power spectrum. This can be attributed to the fact that
the peaks capture information from non-Gaussian features of the
convergence maps. We have shown in Fig.~\ref{fig: pspk} and
Table~\ref{tab: ContourSize} that combining power spectrum and peak
counts improves the constraints by a factor of $\approx$ two,
compared to using the power spectrum alone.

The potential of the peak counts have not yet been fully realized.  
Our work can be improved further by:

(1)  Examining the effects of additional smoothing scales, binning of peaks, and the
 robustness of the results under masking. We have examined only six 
smoothing scales, and demonstrated that using
multiple smoothing scales can reduce the size of the area of the error
contour by a moderate amount. We also showed that masking can change
the power spectrum. More detailed study on these effects can be beneficial.

(2) Including the cosmological dependence of the covariance matrix, 
especially for peak counts. We use a constant covariance matrix in this 
work, assuming the cosmological dependence is weak, as we expect the 
covariance to be dominated by the shape noise. However, as the survey size
increases, cosmological sensitivity should be taken into consideration when
constructing the covariance matrix.

(3) Increasing the number of independent simulations run for each 
cosmological model. In our current work, due
to computational limitations, we have only used one independent N-body
simulation per model. Although we randomly rotate and shift the
lensing planes to create multiple pseudo-independent realizations,
some outliers (such as massive halos) will inevitably be repeated in
several maps. However, our previous work has shown that the bulk of
the cosmological information from peak counts resides in low-amplitude
peaks, which do not arise from single massive halos; these peaks
should be less susceptible to repeated structures between
pseudo-random realizations.  Nevertheless, to test possible errors due
to not having sufficiently independent maps, we ran a separate set of
50 simulations for one cosmology. 

We found that the variance in the 
(noiseless) power spectrum and peak counts is
increased by approximately 10\%, when compared to that using only one
simulation. However, when noise is added, the difference is no longer
systematic, with a 5\% fluctuation and is consistent with 0.
We also found a larger-than Gaussian variance even at our
lowest $\ell = 400$, by approximately 10\%. This increase in the variance
due to non-Gaussianities is somewhat lower than that found previously \cite{Sato2009}. 
Further details on tests of the covariance matrices will be presented in our
companion paper (Petri et al., in prep).

Future WL surveys, such as the Dark Energy Survey, and the Large
Synoptic Survey Telescope, cover much larger areas (5,000 and 20,000
deg$^2$, respectively), hence are more sensitive to instrumental and
theoretical systematics.  These will need to be addressed carefully in
order to realize the full potential of these larger surveys.

\begin{acknowledgments}
  We thank Ludovic Van Waerbeke and Hendrik Hildebrandt for useful discussions. 
  Simulations for this work were performed at the NSF Extreme Science
  and Engineering Discovery Environment (XSEDE), supported by grant
  number ACI-1053575, and at the New York Center for Computational
  Sciences, a cooperative effort between Brookhaven National
  Laboratory and Stony Brook University, supported in part by the
  State of New York. This work was supported in part by the
  U.S. Department of Energy under Contract No. DE-AC02-98CH10886 
  and Contract No. DE-SC0012704, and
  by the NSF grant AST-1210877 (to Z.H.). 

\end{acknowledgments}

\bibliographystyle{physrev}

\end{document}